\documentclass[usenatbib]{mn2e}

\usepackage{graphicx}
\usepackage{mathptmx}
\usepackage[hang,raggedright,nooneline]{subfigure}
\graphicspath{{./Images/}}
\begin{document}
\title[A 610-MHz survey of the ELAIS-N1 field]{A 610-MHz
  survey of the ELAIS-N1 field with the Giant Metrewave Radio
  Telescope -- Observations, data analysis and source catalogue}

\author[T.\ Garn et al.]{Timothy Garn\thanks{E-mail: tsg25@cam.ac.uk},
                        David A.\ Green,
                        Julia M.\ Riley,
                        Paul Alexander\\
                        Astrophysics Group,
                        Cavendish Laboratory, 19 J.~J.~Thomson Ave.,
                        Cambridge CB3~0HE}
\date{\today}
\pubyear{2007}
\maketitle
\label{firstpage}

\begin{abstract}
Observations of the ELAIS-N1 field taken at 610~MHz with the Giant
Metrewave Radio Telescope are presented.  Nineteen pointings were
observed, covering a total area of $\sim9$~deg$^{2}$ with a resolution
of 6~$\times$~5~arcsec$^{2}$, PA $+45\degr$.  Four of the pointings
were deep observations with an rms of $\sim40$~$\mu$Jy before primary
beam correction, with the remaining fifteen pointings having an rms of
$\sim70~\mu$Jy.  The techniques used for data reduction and production
of a mosaicked image of the region are described, and the final mosaic
is presented, along with a catalogue of 2500 sources detected above
6$\sigma$.  This work complements the large amount of optical and
infrared data already available on the region.  We calculate 610-MHz
source counts down to 270~$\mu$Jy, and find further evidence for the
turnover in differential number counts below 1~mJy, previously seen at
both 610~MHz and 1.4~GHz.
\end{abstract}


\section{Introduction}
The {\it Spitzer} Wide-area Infrared Extragalactic
\citep[SWIRE;][]{Lonsdale03} survey has the largest sky coverage of
the legacy surveys being performed by the {\it Spitzer Space
Telescope} \citep{Werner04}.  A total area of $\sim$49~deg$^{2}$ of
sky has been observed with the Infrared Array Camera
\citep[IRAC;][]{Fazio04} and Multiband Imaging Photometer for {\it
Spitzer} \citep[MIPS;][]{Rieke04} instruments at 3.6, 4.5, 5.8, 8, 24,
70 and 160~$\mu$m.  The survey is broken down into six fields, three
in the northern sky -- ELAIS-N1, ELAIS-N2 and the Lockman Hole -- and
three in the south -- ELAIS-S1, {\it Chandra} Deep Field South and the
{\it XMM}-Large Scale Structure (XMM-LSS) field.  All six regions were
selected to be away from the Galactic disk, in order to minimize
background cirrus emission.

There is a large amount of multi-wavelength information available on
all six SWIRE fields.  The three ELAIS fields were observed as part of
the European Large-Area {\it ISO} Survey \citep{Oliver00}, which also
included another northern (-N3) and southern (-S2) field.  The {\it
Infrared Space Observatory} ({\it ISO}) observed these regions at 6.7,
15, 90 and 175~$\mu$m, and a large number of followup observations
were carried out in the optical, infrared and radio bands.  A
band-merged catalogue, containing the {\it ISO} data, along with $U$,
$g'$, $r'$, $i'$, $Z$, $J$, $H$ and $K$-band detections, and radio
observations at 1.4~GHz has been produced -- for more details, see
\citet{RowanRobinson04}, and references therein.

Observations of the ELAIS-N1 region were taken with {\it Spitzer} in
2004 January, covering $\sim9$~deg$^{2}$ with the IRAC and MIPS
instruments.  The source catalogues have been produced, and are
available online \citep{Surace04}, containing over 280,000 sources.
The UK Infrared Deep Sky Survey \citep[UKIDSS;][]{Lawrence07} intends
to cover the ELAIS-N1 region in its Deep Extragalactic Survey plan,
observing the full field in the $J$, $H$ and $K$-bands to a depth of
$K$ = 21~mag.  This will be a great improvement over the currently
available surveys, which have a sensitivity limit of $\sim$18~mag in
the $K$ band.  Data Release 2 \citep{Warren07} of UKIDSS contains
early shallow data on the ELAIS-N1 region.  Further surveys have been
carried out in the $R$-band \citep{Fadda04}, in H$\alpha$
\citep{Pascual01}, and with the {\it Chandra} X-ray telescope
\citep{Manners03,Franceshini05}.  There have been several redshift
surveys of the region \citep{Trichas06,Berta07}, and the ELAIS-N1
region was also partially covered by the Sloan Digital Sky Survey
\citep[SDSS;][]{AdelmanMcCarthy07}.

While there have been a great number of observations of the ELAIS-N1
region at optical and infrared wavelengths, there is comparatively
little radio information available.  The existing VLA 1.4~GHz survey
of the three northern ELAIS fields \citep{Ciliegi99}, which has been
included into the band-merged catalogue of \citet{RowanRobinson04},
reaches a 5$\sigma$ limit of 0.135~mJy over 0.12~deg$^{2}$ but only a
1.15~mJy limit over its full coverage area of 4.22~deg$^{2}$.  The
NVSS \citep{Condon98} and FIRST \citep{Becker95} surveys both cover
the ELAIS-N1 region, but only to relatively shallow $5\sigma$ limits
of 2.25 and 0.75~mJy respectively.  A recent study of polarised
compact sources \citep{Taylor07} at 1420~MHz is underway, using the
Dominion Radio Astrophysical Observatory Synthesis Telescope (DRAO ST)
centered on $16^{\rm h}11^{\rm m}00^{\rm s}$, $+55\degr00'00''$ and
covering 7.4~deg$^{2}$.  The first 30~per~cent of observations have
been completed, with maps in Stokes I, Q and U being produced with a
maximum sensitivity of 78~$\mu$Jy~beam$^{-1}$, although with a
resolution of $\sim$1~arcmin$^{2}$.

In order to extend the information on this region, a much larger deep
radio survey is required.  In this paper, we present observations of
the ELAIS-N1 survey field taken at 610~MHz with the Giant Metrewave
Radio Telescope \citep[GMRT;][]{Ananthakrishnan05}, covering
$\sim9$~deg$^{2}$ of sky with a resolution of $6\times5$~arcsec$^{2}$,
PA~$+45\degr$, centred on $16^{\rm h}11^{\rm m}00^{\rm s}$,
$+55\degr00'00''$ (J2000 coordinates, which are used throughout this
paper).  This survey, in combination with the deep {\it Spitzer} data,
will be used to study the infrared/radio correlation for star-forming
systems \citep[e.g.][]{Appleton04}, and the link between the
triggering of star formation and AGN activity, as well as the
properties of the faint radio population at 610~MHz.

In Section~\ref{sec:observations} we describe the observations and
data reduction techniques used in the creation of the survey.
Section~\ref{sec:results} presents the mosaic and a source catalogue
containing 2500 sources above 6$\sigma$, along with a sample of
extended sources.  In Section~\ref{sec:sourcecounts} we construct the
610~MHz differential source counts, and compare them to previous works.

\section{Observations and data reduction}
\label{sec:observations}
\begin{figure}
  \includegraphics[width=8cm]{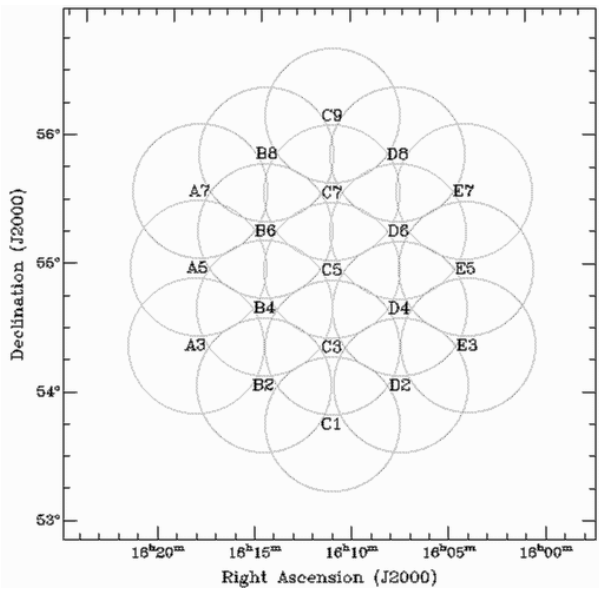}
  \caption{The 19 pointings observed in the ELAIS-N1 region.  Fields
  B4, C3, C5 and D4 are the deep observations.}
  \label{fig:pointings}
\end{figure}

The ELAIS-N1 region was observed for 25~hours, spread over three days
in 2003 August, using the GMRT operating at 610~MHz.  Nineteen
pointings were observed, centred on $16^{\rm h}11^{\rm m}00^{\rm s}$,
$+55\degr00'00''$ and spaced by $36'$ in a hexagonal grid (as shown in
Fig.~\ref{fig:pointings}) in order to get nearly uniform coverage over
the region.  Two 16~MHz sidebands were observed, each split into 128
spectral channels, with a 16.9~s integration time.

The flux density scale was set through observations of 3C48 or 3C286,
at the beginning and end of each observing session.  The {\sc aips}
task {\sc setjy} was used to calculate 610~MHz flux densities of 29.4
and 21.1~Jy, respectively, using the {\sc aips} implementation of the
\citet{Baars77} scale.  Each field was observed for a series of
interleaved 9~min scans in order to maximise the $uv$ coverage, and a
nearby phase calibrator, J1634$+$627, was observed for four minutes
between every three scans to monitor any time-dependent phase and
amplitude fluctuations of the telescope.  The measured phase typically
varied by less than 10~degrees between phase calibrator observations.
Most of the fields were observed for four scans (36 mins in total),
while four fields, B4, C3, C5 and D4, were observed for significantly
longer, in order to give a deeper region within the main survey as
there is deep optical information available on this region from UKIDSS
\citep{Warren07}.  B4 and D4 were observed for 11 scans (99 mins)
while C3 and C5 were observed for 12 scans (108 mins).

\begin{figure}
  \includegraphics[width=8cm]{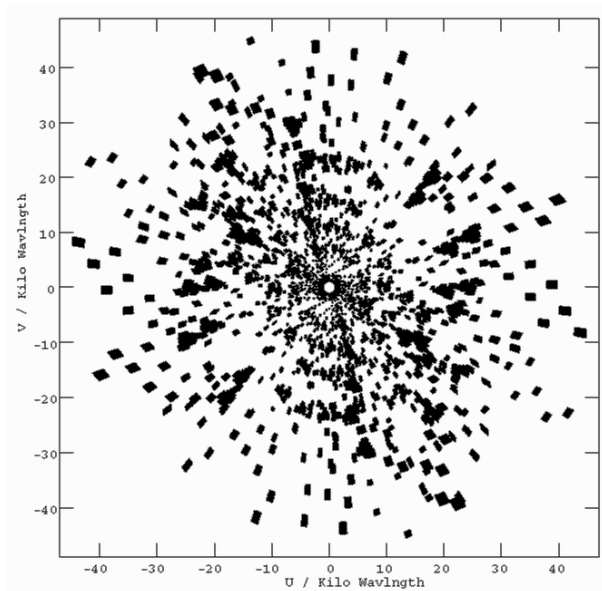}
  \caption{The {\it uv} coverage for pointing A3.  Baselines less than
  1~k$\lambda$ were not used in imaging and have been omitted from the
  figure.}
  \label{fig:uvcoverage}
\end{figure}

The imaging strategy was very similar to that used for our 610~MHz
GMRT observations of the {\it Spitzer} extragalactic First Look Survey
field \citep[xFLS;][]{Garn07}.  An error in the timestamps of the {\it
uv} data was corrected using {\sc uvfxt} (\citealp[for further details
see][]{Garn07}).  This error is only found for observations made
before the summer of 2006, and has since been fixed at the GMRT.
Initial editing of the data was performed separately on each sideband
with standard {\sc aips} tasks, to remove bad baselines, antennas, and
channels that were suffering from large amounts of narrow band
interference, along with the first and last integration periods of
each scan.  The flux calibrators were used to create a bandpass
correction for each antenna.  In order to create a continuum channel,
five central frequency channels were combined together, and an
antenna-based phase and amplitude calibration created using
observations of J1634$+$627.  This calibration was applied to the
original data, which was then compressed into 11 channels, each with
bandwidth of 1.25~MHz (so the first few and last few spectral channels
were omitted from the data, since they tended to be the noisiest).
The new channel width is small enough that bandwidth smearing is not a
problem in our images, and led to an effective bandwidth of 13.75~MHz
in each sideband.  Further flagging was performed on the 11 channel
data set, and the two sidebands combined using {\sc uvflp}
\citep[again, see][]{Garn07} to improve the $uv$ coverage.  The
coverage for one of the shallow observations is shown in
Fig.~\ref{fig:uvcoverage}.  Baselines shorter than 1~k$\lambda$ were
omitted from the imaging, since the GMRT has a large number of small
baselines which would otherwise dominate the beam shape.

Each pointing was broken down into 31 smaller facets \citep[as
discussed in][]{Garn07}, arranged in a hexagonal grid and covering an
area with diameter $\sim1\fdg8$.  These were imaged separately, each
with a different assumed phase centre.  The large area covered
(compared with the Full-Width Half-Maximum of the GMRT, which is
$\sim0\fdg74$) allows bright sources well outside of the observed
region to be cleaned from the images, while the faceting procedure
avoids the introduction of phase errors due to the non-planar nature
of the sky.  All images were made with the same elliptical synthesised
beam of size $6 \times 5$~arcsec$^{2}$, position angle $+45\degr$ by
setting the parameters for the restoring beam within {\sc imagr}.  A
pixel size of 1.5~arcsec was chosen, to ensure that the beam was well
oversampled.

Each pointing went through three iterations of phase self-calibration
at 10, 3 and 1 minute intervals, and then a final round of
self-calibration correcting both phase and amplitude errors, at
10~minute intervals.  The overall amplitude gain was held constant in
order not to alter the flux density of sources.  The self-calibration
steps improved the noise level by about 10~per~cent, and significantly
reduced the residual sidelobes around the brighter sources.

The four deep pointings have a final rms of between 40 and
42~$\mu$Jy before correction for the GMRT primary beam, while the 15
shallow fields have a noise level of between 66 and 73~$\mu$Jy.  These
figures are very close to the expected thermal noise limits of 36 and
63~$\mu$Jy respectively, calculated using 

\begin{equation}
  \sigma = \frac{\sqrt{2}T_{\rm s}}{G\sqrt{n(n-1)N_{\rm IF}
  \Delta\nu\tau}}
  \label{eq:noise}
\end{equation}
where the system temperature $T_{\rm s} \approx$ 92~K, and the antenna gain $G
\approx$ 0.32~K~Jy$^{-1}$ -- values taken from the GMRT
website\footnote{\tt {\scriptsize
  http://www.gmrt.ncra.tifr.res.in/gmrt\_hpage/Users/Help/help.html}}
-- $n$ is the number of working antennas, typically 27 during our
observations, $N_{\rm IF} = 2$ is the number of sidebands, $\Delta\nu$
= 13.75~MHz is the effective bandwidth per sideband, and $\tau$ is the
integration time for each pointing.

\begin{figure}
  \includegraphics[width=8cm]{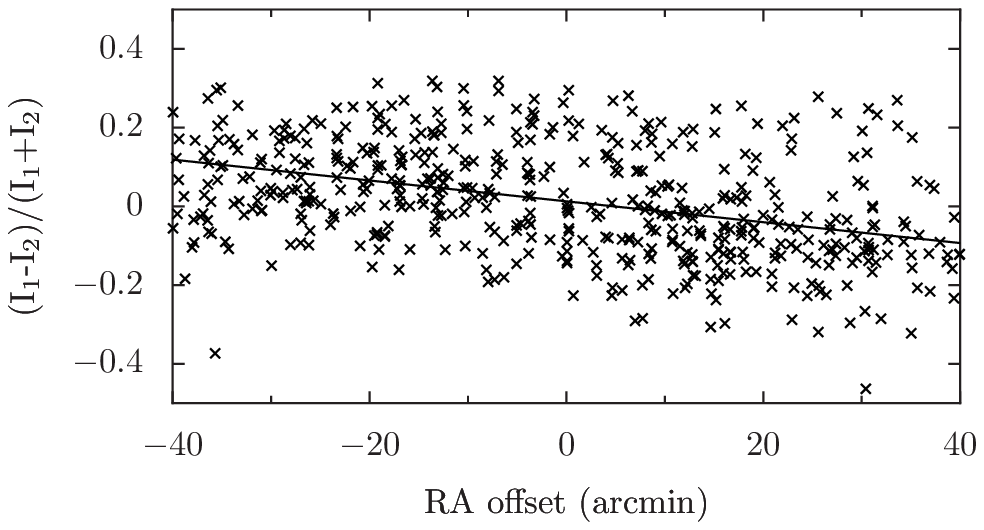}
  \caption{Fractional offset in peak brightness of all sources in
  overlapping regions, using the nominal primary beam centre.  A
  gradient across the pointing is seen, leading to systematic errors
  in measured flux density.}
  \label{fig:offsetbefore}
\end{figure}

In \citet{Garn07}, we detected a position-dependent error with the
GMRT primary beam.  This led to a systematic difference between the
measured flux densities of sources that were present in more than one
pointing, with the fractional offset varying across the primary beam.
To check for a similar effect here, we corrected each pointing for the
nominal beam shape, using an 8th-order polynomial with coefficients
taken from \citet{Kantharia01}.  We took the 557 sources present in
the overlapping regions between two pointings with peak brightness
greater than 1.5~mJy~beam$^{-1}$, and looked at the fractional offsets
as a function of position away from each pointing centre.
Fig.~\ref{fig:offsetbefore} shows the variation in this offset with
source Right Ascension.  A similar trend was seen with Declination.

\begin{figure}
  \includegraphics[width=8cm]{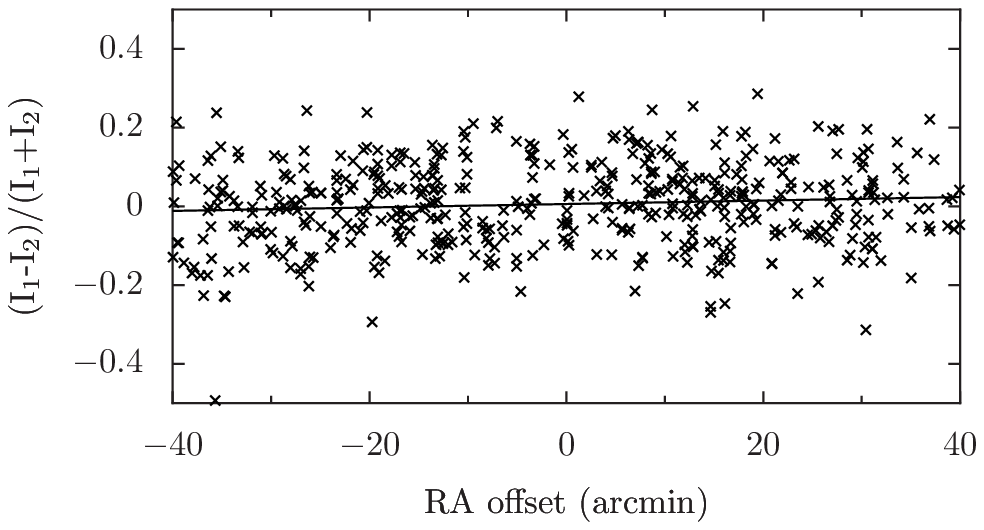}
  \caption{Fractional offset in peak brightness of sources in
  overlapping regions, using the shifted primary beam centre applied
  to all fields.}
  \label{fig:offsetafter}
\end{figure}

We repeated the analysis of \citet{Garn07}, to model the effective
pointing centre of the telescope as having a systematic offset from
its nominal value.  By shifting the phase centre of all pointings by
$\sim2\farcm7$ in a north-west direction before performing the primary
beam correction, we were able to remove this systematic effect, as
shown in Fig~\ref{fig:offsetafter}.  The amount and direction of the
correction was consistent between pointings, and did not vary between
the deep and shallow observations.  This correction also removed the
systematic effect seen in the Declination direction.  The size and
direction of this correction is similar to that seen in the xFLS
survey.

\begin{figure}
  \includegraphics[width=8cm]{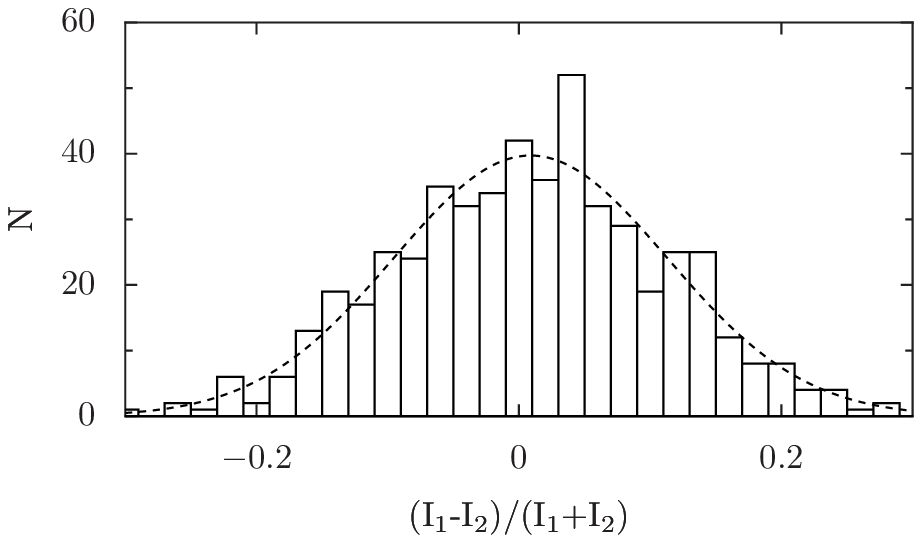}
  \caption{Distribution of fractional offsets, after correction
  for the shifted primary beam.  The best-fit Gaussian is shown,
  giving a 10~per~cent error in the overall flux density calibration.}
  \label{fig:beamaccuracy}
\end{figure}

The accuracy of the primary beam correction was then tested, as shown
in Fig.~\ref{fig:beamaccuracy}.  The fractional offset of sources in
overlapping regions is now a good fit to a Gaussian, with mean of
0.009 and $\sigma$ of 0.1, indicating a 10~per~cent error in the
absolute flux density calibration of all sources.

The 19 pointings were mosaicked together, taking into account the
offset primary beam and weighting the final mosaic appropriately by the
relative noise of each pointing.  The mosaic was cut off at the point
where the primary beam correction dropped to 20~per~cent of its
central value.  The final mosaicked image is available via {\tt
http://www.mrao.cam.ac.uk/surveys/}.

\section{Results and discussion}
\label{sec:results}
\begin{figure}
  \includegraphics[width=8cm]{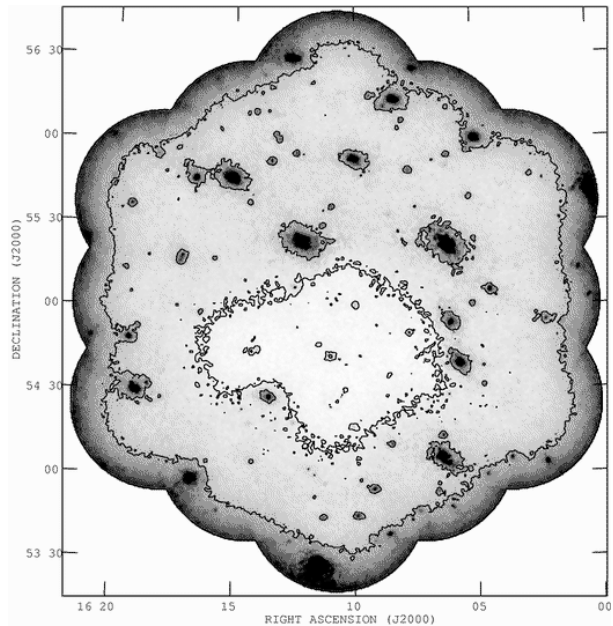}
  \caption{The rms noise of the final mosaic, calculated using Source
  Extractor.  The grey-scale ranges between 40 and 350~$\mu$Jy, and
  the contours are at 60 and 120~$\mu$Jy respectively.}
  \label{fig:EN1noise}
\end{figure}

Source Extractor \citep[SExtractor;][]{Bertin96} was used to calculate
the rms noise $\sigma$ across the mosaic.  A grid of $16\times16$
pixels was used in order to track changes in the local noise level,
which varies significantly near the brightest sources.
Fig.~\ref{fig:EN1noise} illustrates the local noise, with the
grey-scale varying between 40~$\mu$Jy (the noise level in the centre
of the deep pointings) and 350~$\mu$Jy (the noise level for the
shallow pointings, at the distance where the GMRT primary beam gain
was 20~per~cent of its central value).  The 60 and 120~$\mu$m contours
are plotted, which cover the majority of the deep and shallow survey
regions respectively.

\begin{figure*}
  \includegraphics[width=16cm]{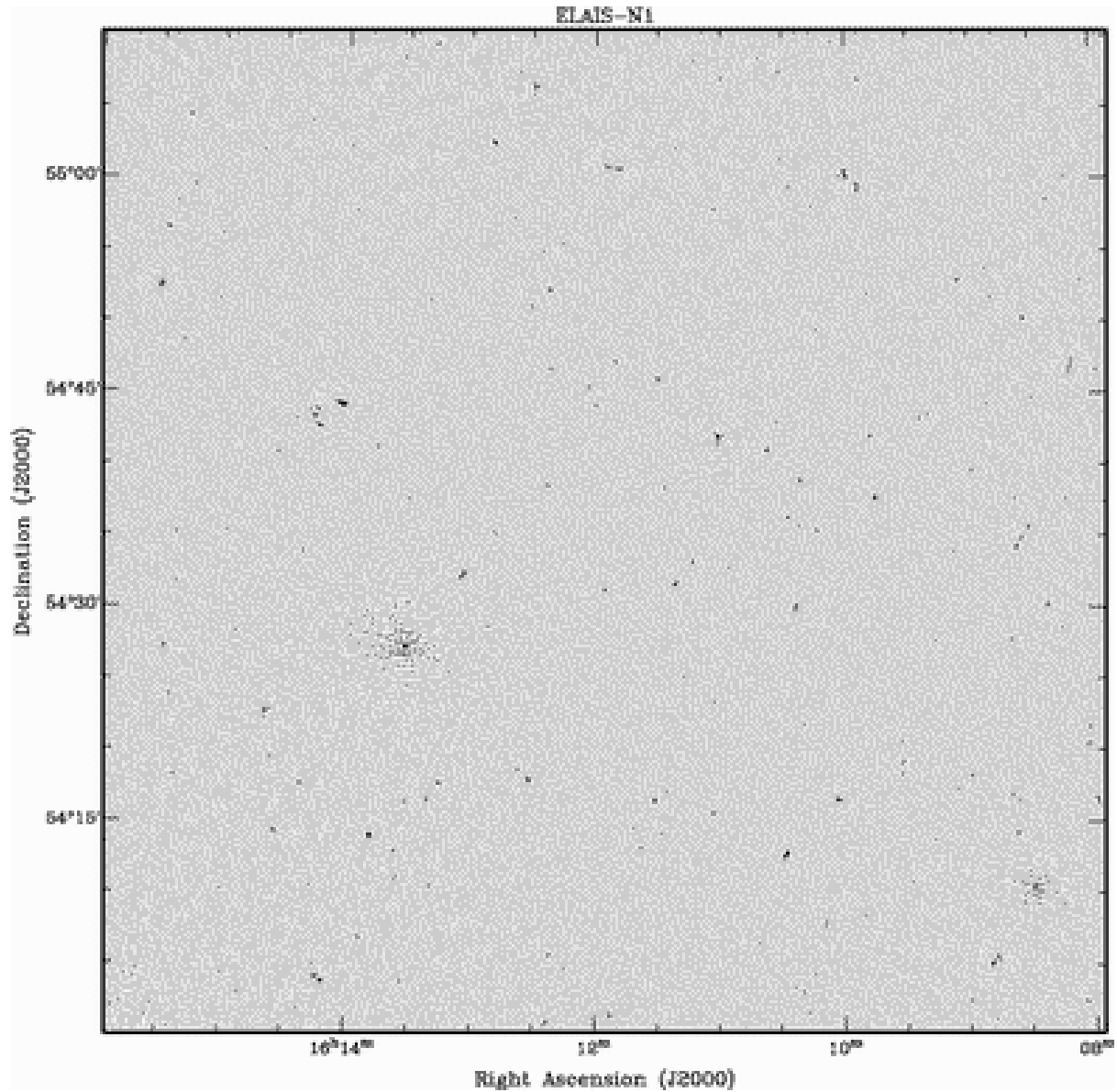}
  \caption{A $70\times70$~arcmin$^{2}$ region of the 610~MHz image, to
  illustrate the quality of the survey.  The region is located within
  the deeper area of the survey, and most sources are unresolved. The
  grey-scale ranges between $-0.2$ and 1~mJy~beam$^{-1}$, and the
  noise is relatively uniform and between 40 and 60~$\mu$Jy, apart
  from small regions near bright sources where the noise increases.}
  \label{fig:EN1sample}
\end{figure*}

\begin{figure*}
  \centerline{\subfigure[GMRTEN1~J161137.8$+$555955]
                {\includegraphics[width=5cm]{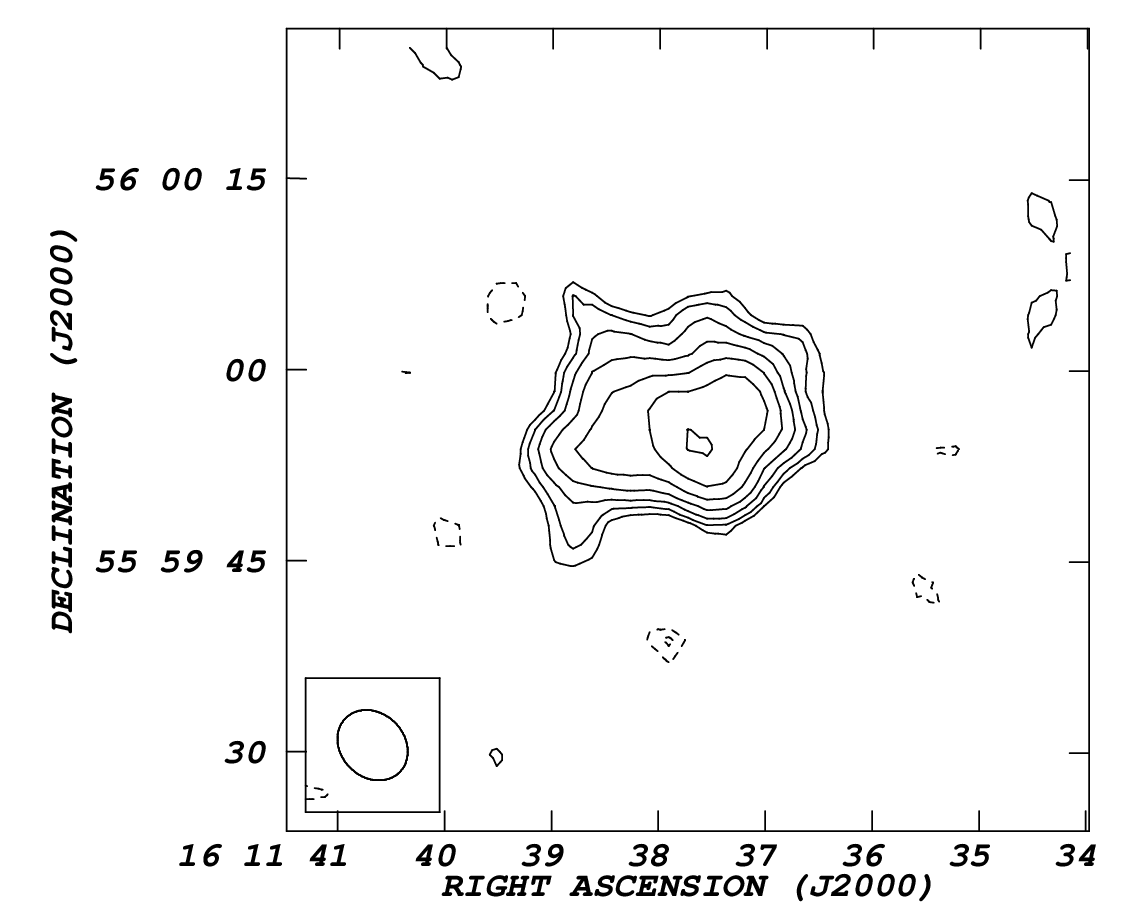}}
              \subfigure[GMRTEN1~J160640.9$+$560136]
                {\includegraphics[width=5cm]{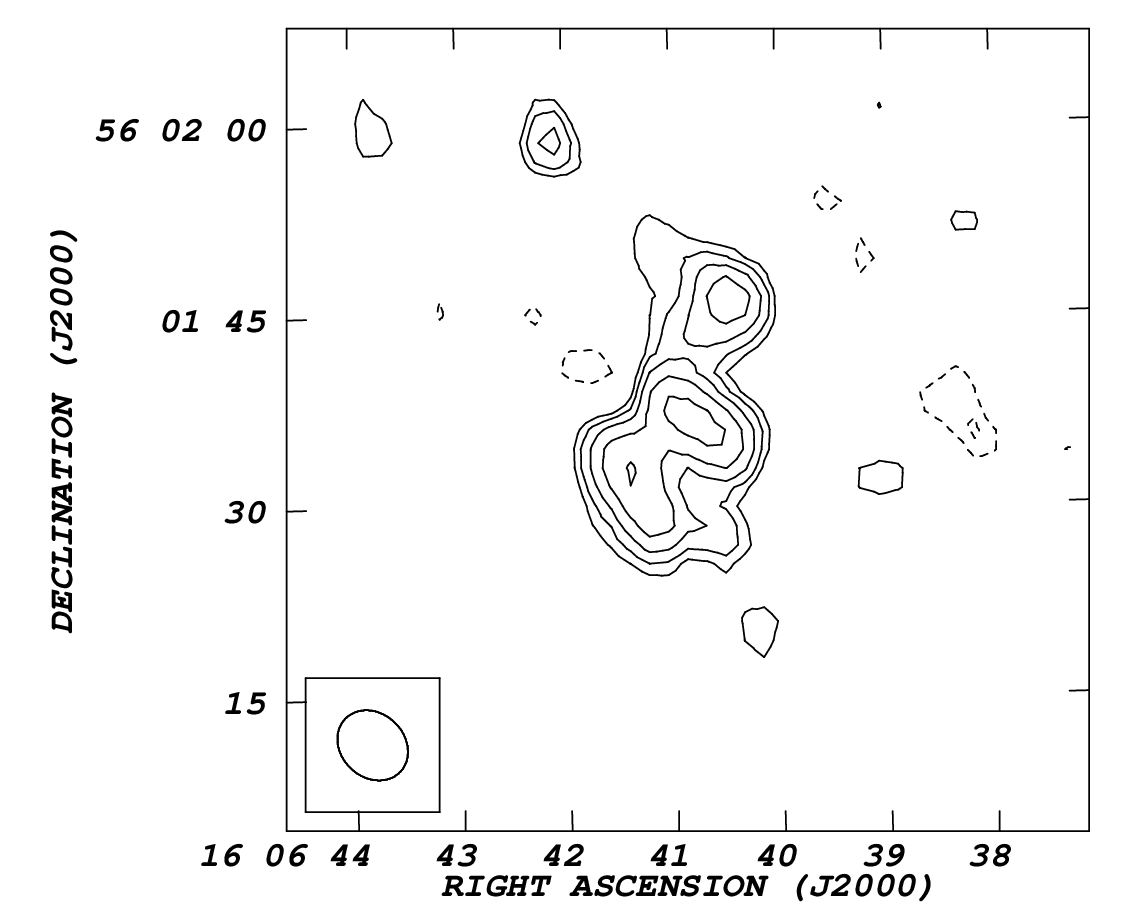}}
              \subfigure[GMRTEN1~J161530.7$+$545231]
                {\includegraphics[width=5cm]{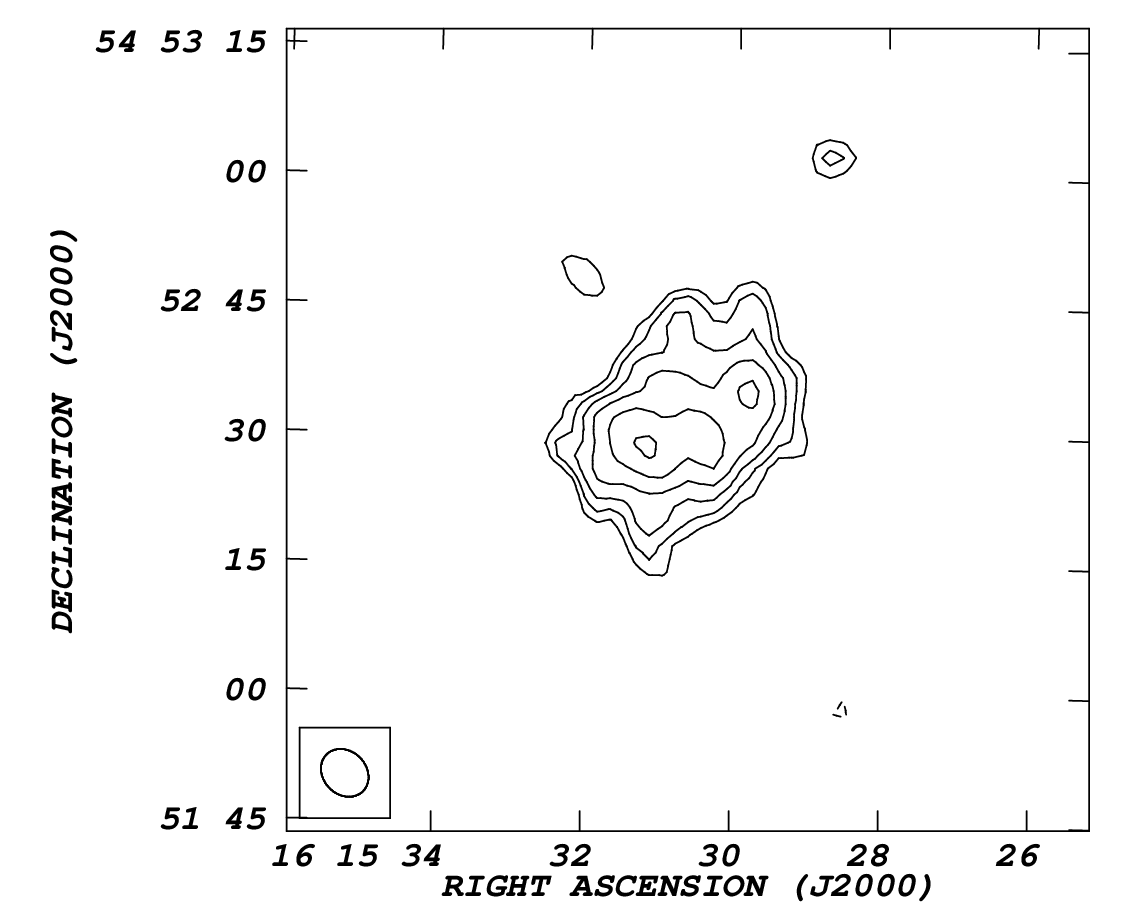}}}
  \centerline{\subfigure[GMRTEN1~J160929.9$+$552444 and
               J160931.09$+$552503]
                 {\includegraphics[width=5cm]{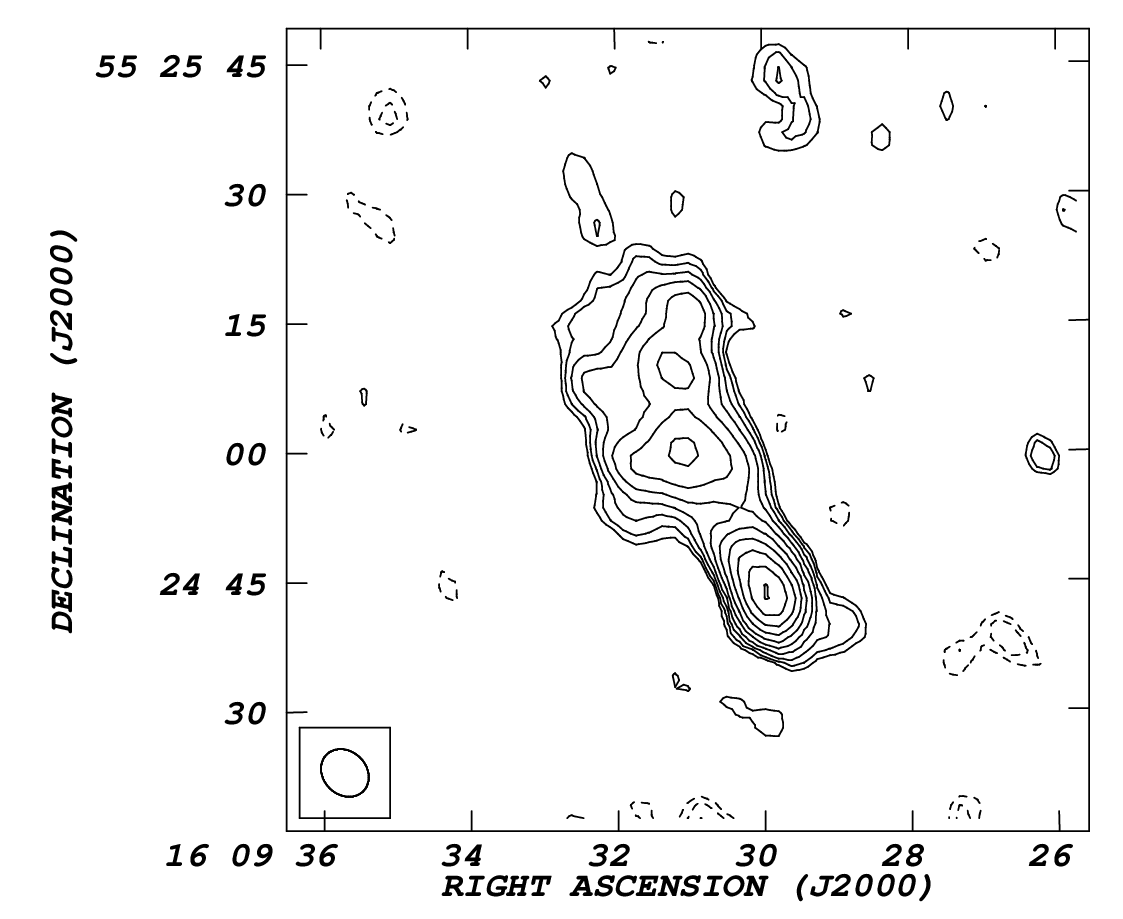}}
              \subfigure[GMRTEN1~J161858.8$+$545227,
              J161900.3$+$545305 and J161903.3$+$545240]
                 {\includegraphics[width=5cm]{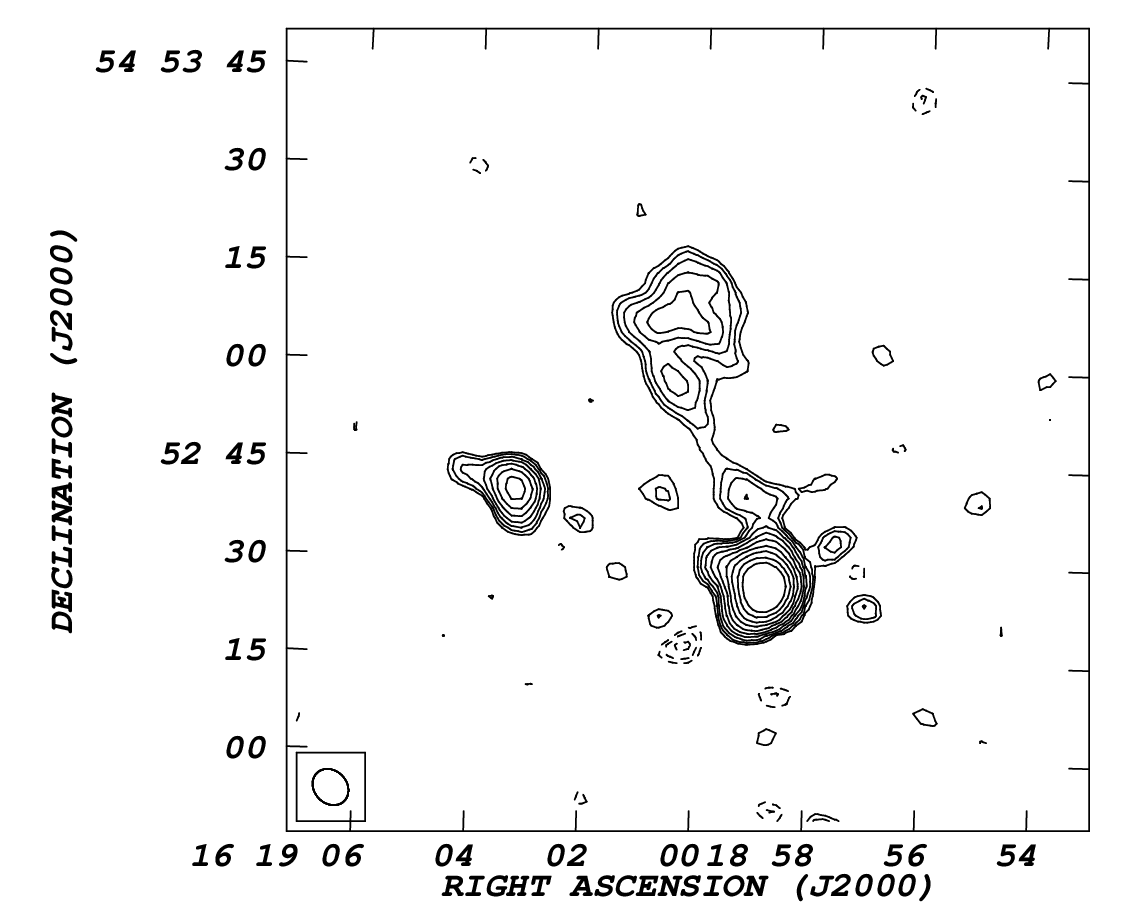}}
              \subfigure[GMRTEN1~J161148.4$+$550049,
              J161151.5$+$550053 and J161154.4$+$550057]
                 {\includegraphics[width=5cm]{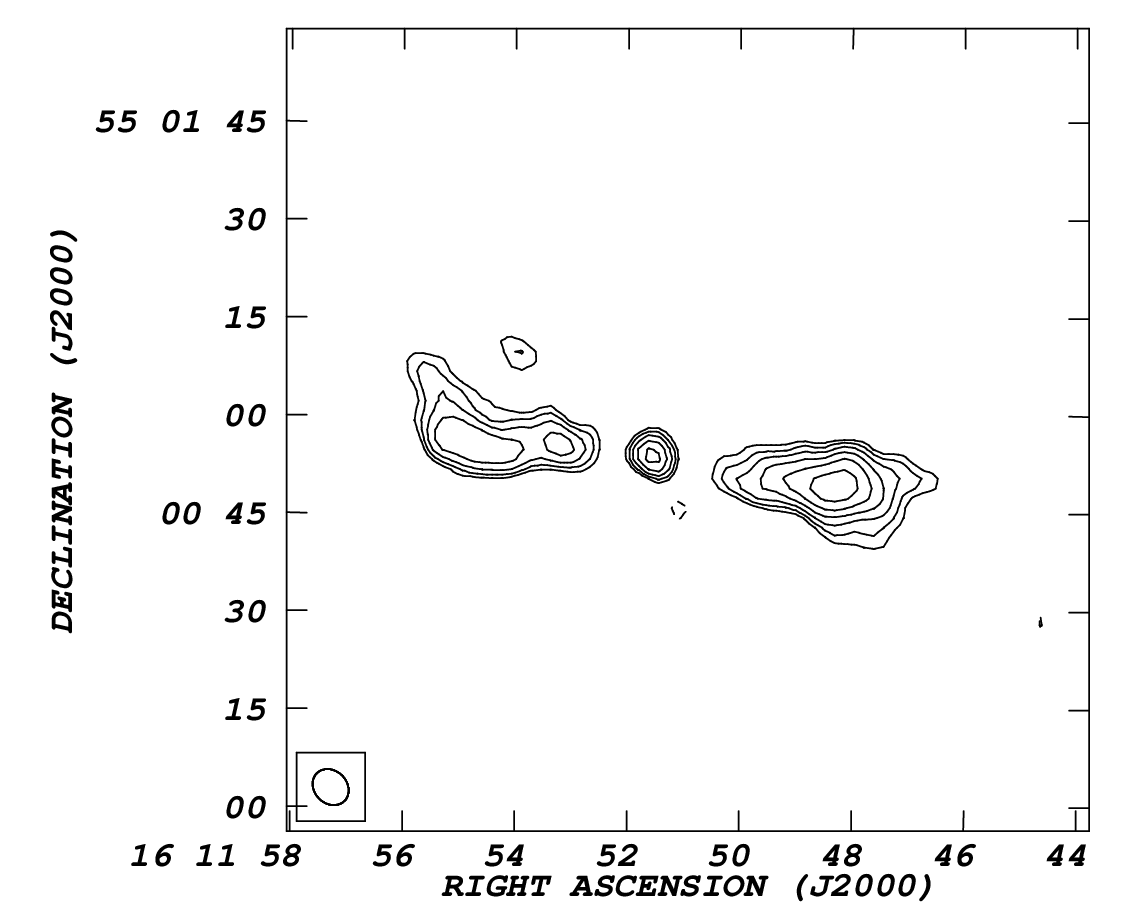}}}
  \centerline{\subfigure[GMRTEN1~J160808.7$+$544723,
              J160809.5$+$544659, J160810.3$+$544633 and 
              J160810.7$+$544645]
                 {\includegraphics[width=5cm]{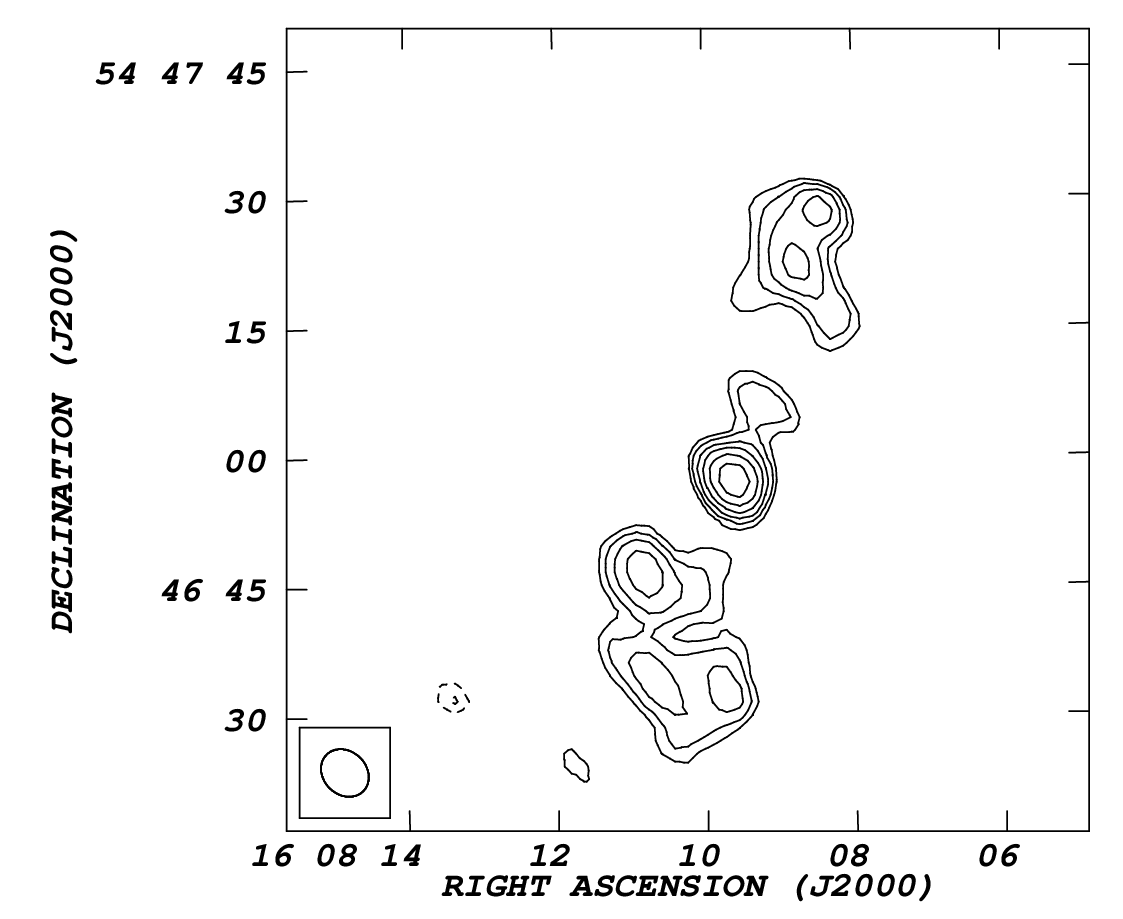}}
              \subfigure[GMRTEN1~J161027.4$+$541246]
                 {\includegraphics[width=5cm]{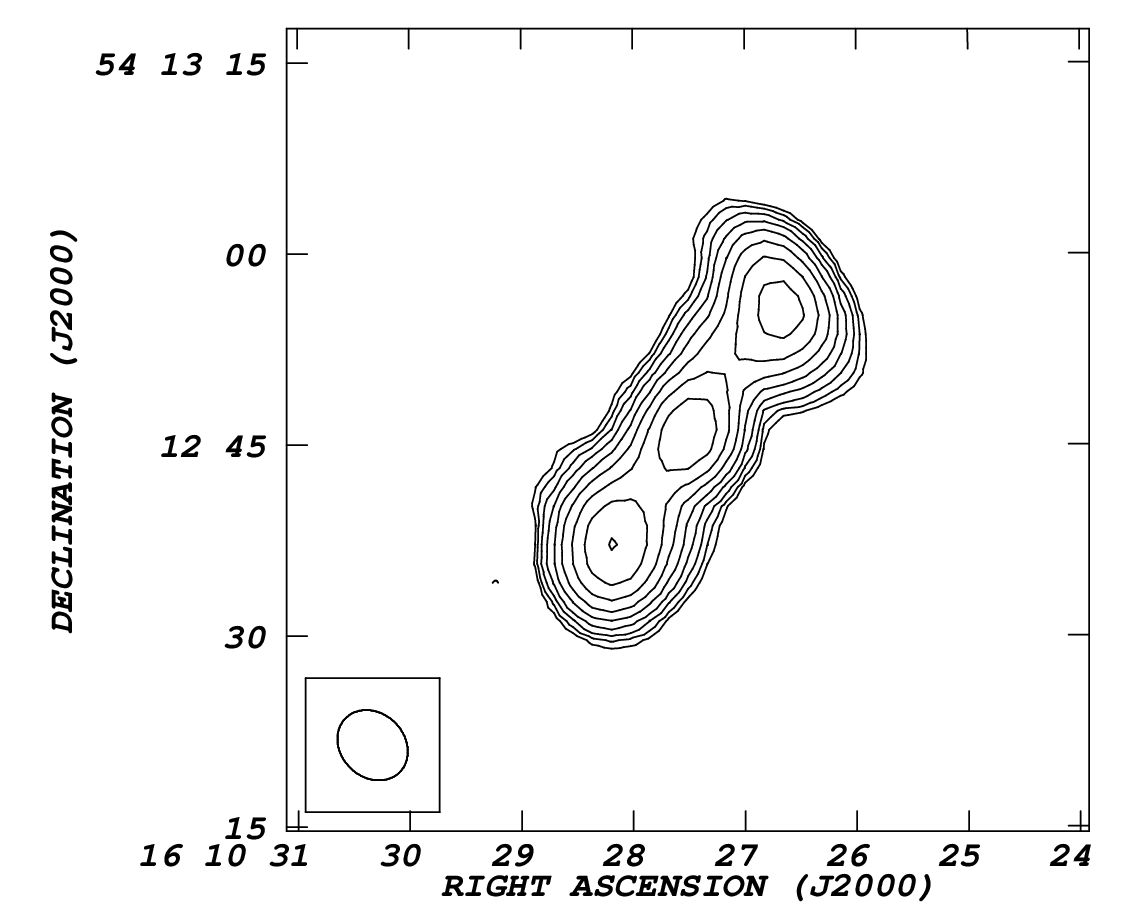}}
              \subfigure[GMRTEN1~J161757.1$+$545110]
                 {\includegraphics[width=5cm]{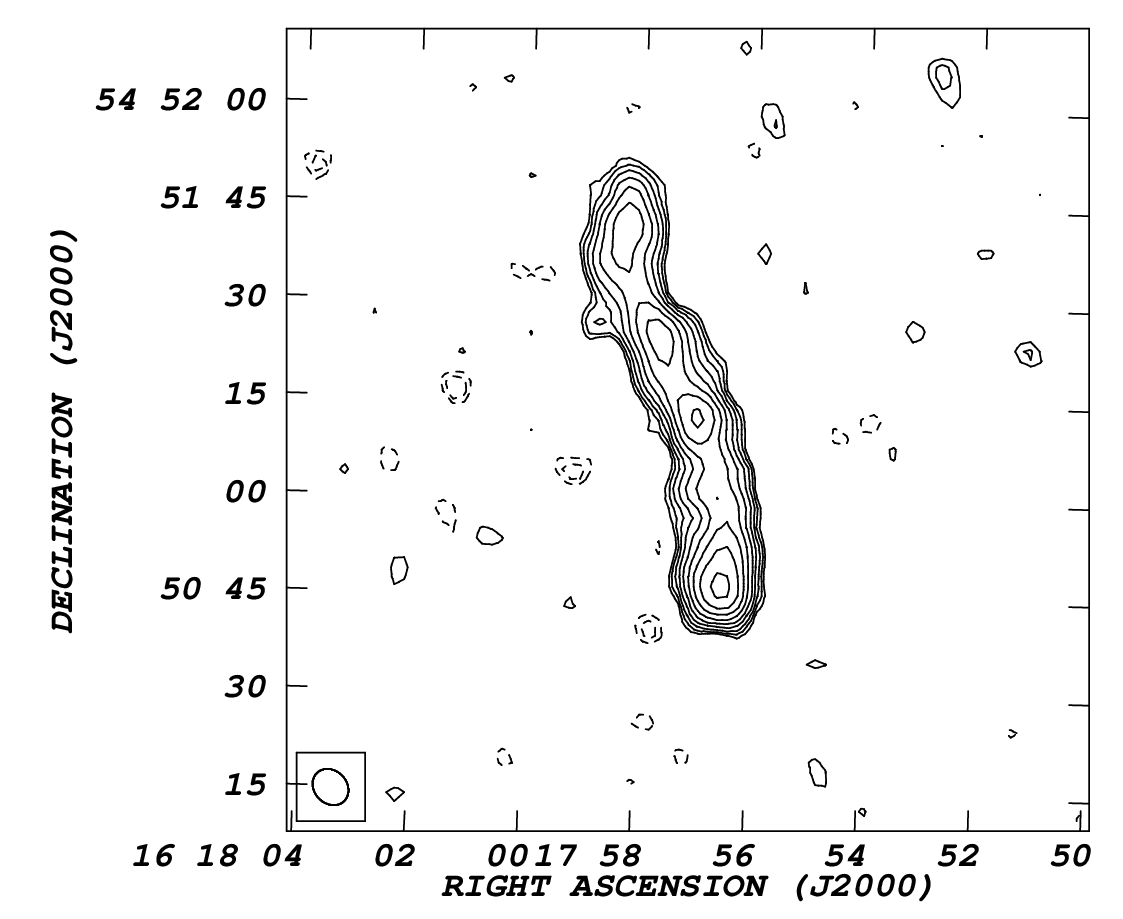}}}
  \centerline{\subfigure[GMRTEN1~J161140.5$+$554703 and 
              J161142.8$+$554727]
                 {\includegraphics[width=5cm]{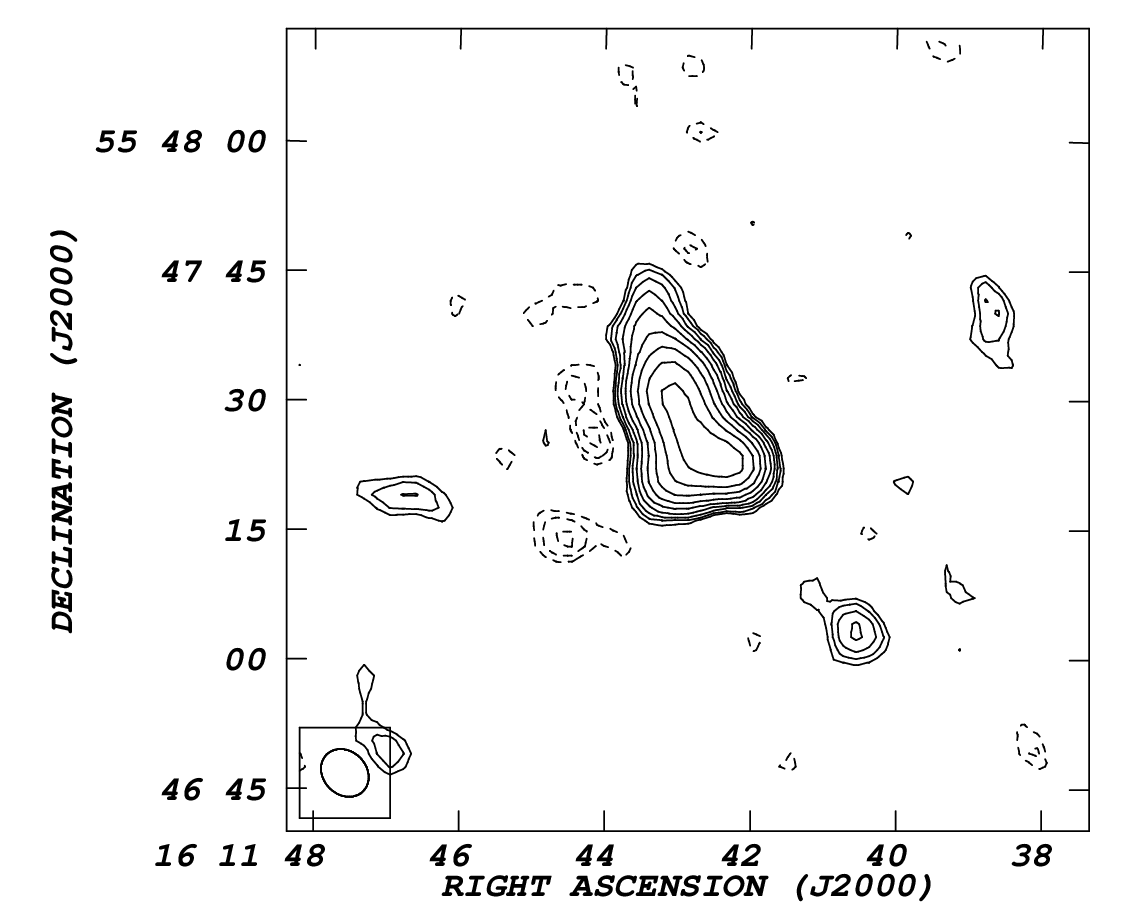}}
              \subfigure[GMRTEN1~J160634.8$+$543456]
                 {\includegraphics[width=5cm]{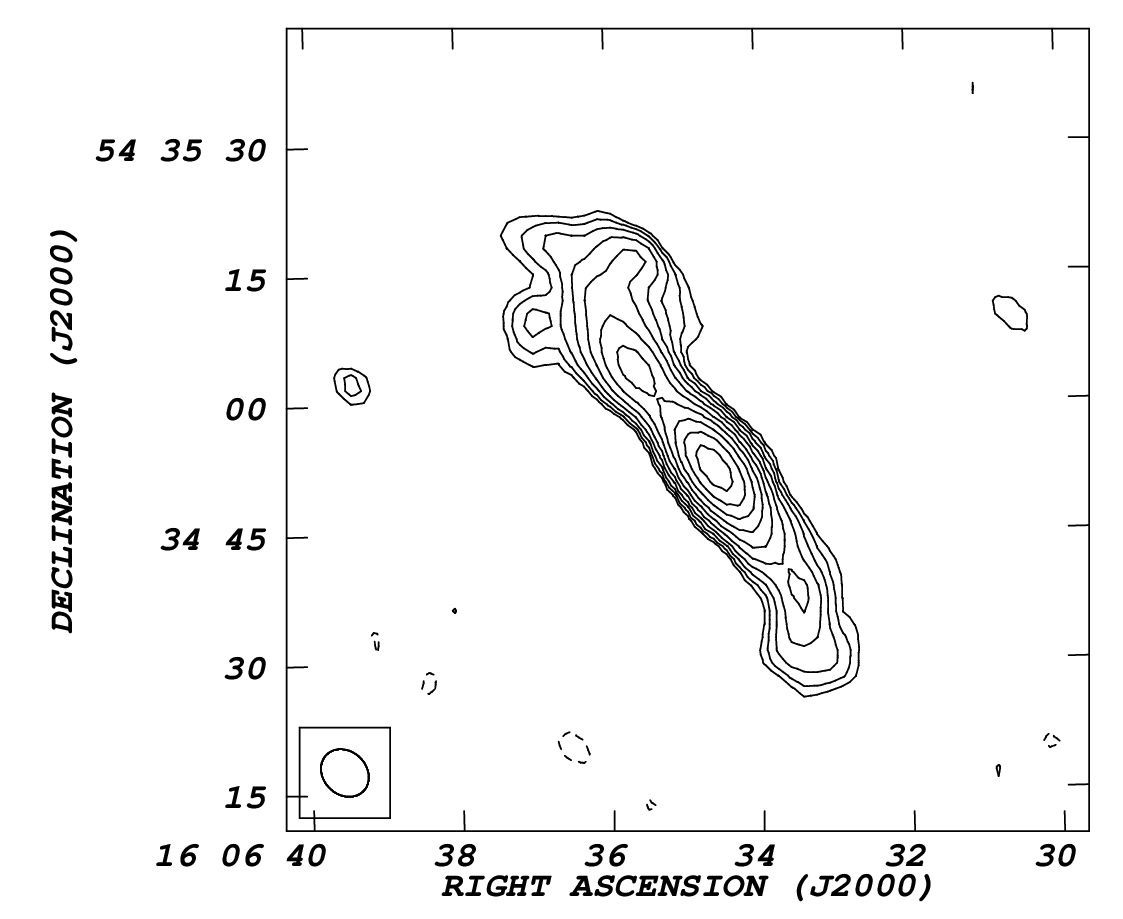}}
              \subfigure[GMRTEN1~J160334.6$+$542900 and 
              J160333.1$+$542914]
                 {\includegraphics[width=5cm]{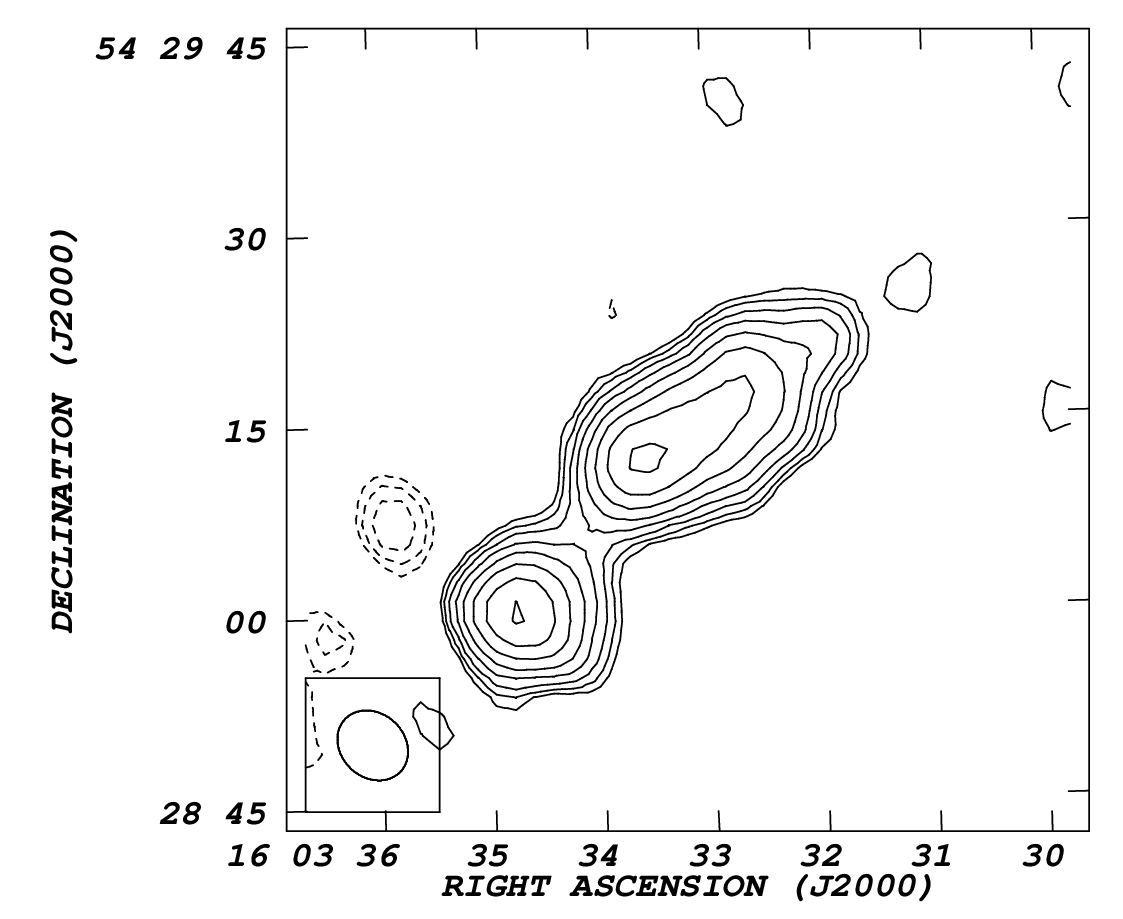}}}
  \caption{A selection of extended objects in the ELAIS-N1 610~MHz
  GMRT survey -- contours are plotted at $\pm200~\mu$Jy $\times$ 1,
  $\sqrt{2}$, 2, $2\sqrt{2}$, 4$\ldots$, with the exception of object
  (e), where the contours start at $\pm$500~$\mu$Jy.  Negative
  contours are represented by dashed lines.  The resolution of the
  beam is shown in the bottom left of each image, and the designations
  of each source component are given below.}
  \label{fig:extended}
\end{figure*}

A sample region of the ELAIS-N1 survey is shown in
Fig.~\ref{fig:EN1sample} to illustrate the quality of the image.  Most
of the sources in our survey are unresolved, although there are
several with extended structures.  We present a sample of these in
Fig.~\ref{fig:extended}.

\begin{figure}
  \includegraphics[width=8cm]{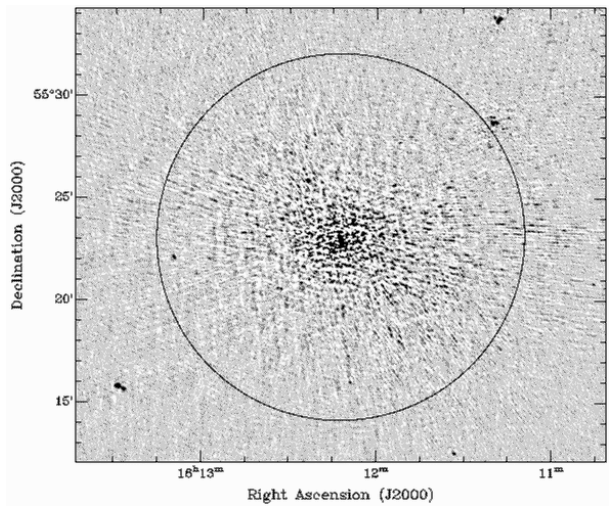}
  \caption{Source GMRTEN1~J161212.4$+$552308, and the errors
  surrounding it.  The grey-scale ranges between $-0.2$ and
  1~mJy~beam$^{-1}$, and the source has a peak brightness of 389~mJy.  The
  region affected by an overdensity of sources is shown by the black
  circle, of radius 10~arcmin -- see text for more details.}
  \label{fig:BrightSource}
\end{figure}

Our GMRT data suffer from dynamic range problems near the brightest
sources, and the final mosaic has increased noise and residual
sidelobes in these regions.  We had fewer problems with our survey of
the xFLS region, due to the longer time spent on each pointing and the
correspondingly better {\it uv} coverage.  There were also fewer
bright sources in the xFLS field, so a much smaller region was
affected by residual sidelobes.  Fig.~\ref{fig:BrightSource} shows an
area around one of the bright sources, to illustrate the problems
caused by the residual sidelobes.  While the local noise calculated by
SExtractor increases due to these residuals, some of them still have
an apparent signal-to-noise level that is greater than 6.  We
therefore opted for a two-stage selection criteria for our final
catalogue.

\subsection{Source fitting}
An initial catalogue of 4767 sources was created using SExtractor.
The mosaic was cut off at the point where the primary beam correction
dropped to 20~per~cent of its central value (a radius of 0\fdg53 from
the outer pointings), however only sources inside the 30~per~cent
region (0\fdg47) were included in the catalogue to avoid the mosaic
edges from affecting the estimation of local noise.  The requirements
for a source to be included were that it had at least 5 connected
pixels with brightness greater than 3$\sigma$, and a peak brightness
greater than 6$\sigma$.  The image pixel size meant that the beam was
reasonably oversampled, so the source peak was taken to be the value
of the brightest pixel within a source.  The integrated flux density
was calculated using the FLUX\_AUTO option within SExtractor.  This
creates an elliptical aperture around each object \citep[as described
in][]{Kron80}, and integrates the flux contained within the ellipse.
Comparisons between the flux density obtained through this method and
through the method developed in \citet{Garn07} -- pixels above a given
threshold were summed, then empirically corrected for the elliptical
beam shape -- give good agreement between the two techniques.  Sources
with Kron flux density above 1~mJy showed no statistical difference
between the two flux density measurements, with an uncertainty of
3~per~cent.  Sources below 1~mJy had a Kron flux density that was
systematically larger than the \citet{Garn07} method by 6~per~cent,
with an uncertainty of 4~per~cent.  We chose to use the Kron method in
order to avoid the empirical correction factor.

In order to estimate the area affected by artefacts near the bright
sources, we calculated the number of (potentially spurious) sources in
a series of concentric rings centred on a bright source, and converted
this to an effective source density as a function of distance.  While
the noise does vary across the map, the variation is smooth and
relatively slow away from the bright sources.  Using the more distant
rings we calculated a mean source sky density $\mu$, and an estimate
of the error in this value, $\sigma_{\mu}$, which will be unaffected
by the presence of the spurious sources, and will vary slightly
between each bright central source due to the changing properties of
the image.

\begin{figure}
  \includegraphics[width=8cm]{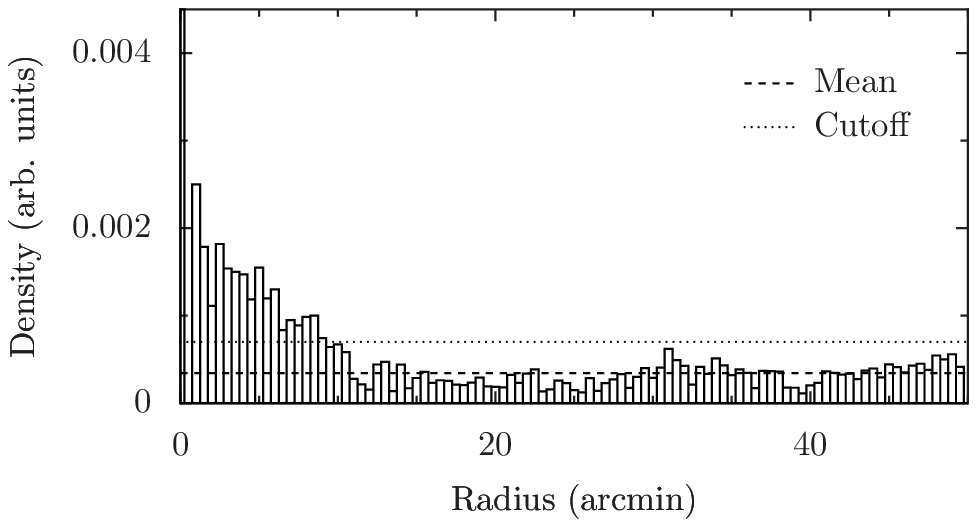}
  \caption{Density of sources, in concentric rings around the
  bright source shown in Fig.~\ref{fig:BrightSource}.  The overdensity of
  sources near to the bright (389~mJy) central object can be clearly
  seen.  The mean source density, far from the central source, is
  given by the large dashed line while the cutoff density defining the
  affected region is given by short dashes -- see text for more details.}
  \label{fig:spurioussources}
\end{figure}

\begin{figure}
  \includegraphics[width=8cm]{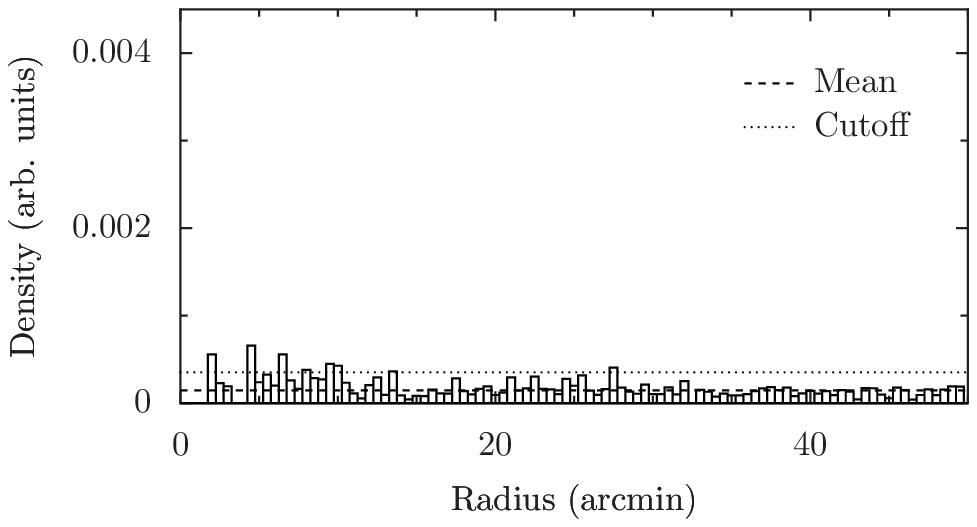}
  \caption{Density of sources, in concentric rings around a fainter
  (10~mJy) object in the ELAIS-N1 survey field.  There is still a
  slight overdensity near the central source, but the signal-to-noise
  cutoff of $6\sigma$ ensures that spurious sources are not included
  in the catalogue.}
  \label{fig:FaintSource}
\end{figure}

Fig.~\ref{fig:spurioussources} shows the source density (in arbitrary
units) away from the source in Fig.~\ref{fig:BrightSource}.  The clear
over-density of sources can be seen within 10~arcmin of the source.
If there is a significant peak in the density (greater than $\mu +
6\sigma_{\mu}$), then we define the size of the affected region by
finding the first radius at which the source density drops below $\mu
+ 3\sigma_{\mu}$ -- for this source, the affected radius is 10~arcmin.
The affected region has been added to Fig.~\ref{fig:BrightSource}.
Within this region, only sources with a peak brightness greater than
12$\sigma$ are included in our catalogue -- this value was determined
empirically.  The source density plot for a 10~mJy source is shown in
Fig.~\ref{fig:FaintSource}, on the same scale as
Fig.~\ref{fig:spurioussources} -- while there may still be a slight
overdensity near the centre, the increased noise level in this region,
along with the $6\sigma$ cut-off reduces the risk of selecting a
spurious source near sources with weak over-densities.

This analysis was repeated for all sources with a peak greater than
10~mJy in order to filter spurious sources.  The final catalogue
contains 2500 sources -- we have erred on the side of caution in order
to produce a catalogue with little contamination from spurious
sources.  The size of the affected region is correlated with the peak
brightness of a source (with Pearson product-moment correlation
coefficient of 0.53), and the number of spurious sources is also
correlated with the peak brightness, with correlation coefficient
0.73.  The precise size of the affected region depends on the {\it
uv} coverage for the relevant pointing, the time spent on observations
and the local noise levels.

Table $\ref{tab:catalogue}$ presents a sample of 60 entries in the
catalogue, which is sorted by right ascension.  The full table is
available via {\tt http://www.mrao.cam.ac.uk/surveys/}.  Column~1
gives the IAU designation of the source, in the form
GMRTEN1~Jhhmmss.s$+$ddmmss, where J represents J2000.0 coordinates,
hhmmss.s represents right ascension in hours, minutes and truncated
tenths of seconds, and ddmmss represents the declination in degrees,
arcminutes and truncated arcseconds.  Columns~2 and 3 give the right
ascension and declination of the source, calculated by first moments
of the relevant pixel brightnesses to give a centroid position.
Column~4 gives the brightness of the peak pixel in each source, in
mJy~beam$^{-1}$, and column~5 gives the local rms noise in
$\mu$Jy~beam$^{-1}$.  Columns~6 and 7 give the integrated flux density
and error, calculated from the local noise level and source size.
Columns 8 and 9 give the $X$, $Y$ pixel coordinates of the source
centroid from the mosaic image.  Column 10 is the Source Extractor
deblended object flag -- 1 where a nearby bright source may be
affecting the calculated flux, 2 where a source has been deblended
into two or more components from a single initial island of flux, and
3 when both of the above criteria apply.  There are 232 sources
present in our catalogue with non-zero deblend flags; it is
necessary to examine the images to distinguish between the case where
one extended object has been represented by two or more entries, and
where two astronomically distinct objects are present.

\begin{table*}
  \label{tab:catalogue}
  \caption{A sample of 60 entries from the 610-MHz ELAIS-N1 catalogue,
  sorted by right ascension.  The full version of this table is
  available as Supplementary Material through the online version of
  this article, and via {\tt http://www.mrao.cam.ac.uk/surveys/}.}
  \begin{tabular}{cccccccccc} 
\hline
Name & RA & Dec.\ & Peak & Local~Noise & Int.\ Flux Density & Error & $X$ & $Y$ & Flags\\
 & J2000.0 & J2000.0 & mJy~beam$^{-1}$ & $\mu$Jy~beam$^{-1}$ & mJy & mJy & & & \\
(1) & (2) & (3) & (4) & (5) & (6) & (7) & (8) & (9) & (10) \\
\hline
GMRTEN1~J160319.2$+$542543 & 16:03:19.22 & $+$54:25:43.6 & 0.496 & 75 & 0.415 & 0.067 & 6599 & 2946 & 0\\
GMRTEN1~J160319.2$+$553149 & 16:03:19.24 & $+$55:31:49.1 & 0.469 & 66 & 0.383 & 0.054 & 6527 & 5589 & 0\\
GMRTEN1~J160319.3$+$554950 & 16:03:19.38 & $+$55:49:50.2 & 0.514 & 85 & 0.265 & 0.054 & 6506 & 6309 & 0\\
GMRTEN1~J160320.8$+$550645 & 16:03:20.89 & $+$55:06:45.9 & 2.617 & 97 & 3.425 & 0.175 & 6545 & 4587 & 0\\
GMRTEN1~J160321.1$+$553654 & 16:03:21.13 & $+$55:36:54.2 & 0.418 & 69 & 0.309 & 0.076 & 6510 & 5792 & 0\\
GMRTEN1~J160322.6$+$554320 & 16:03:22.67 & $+$55:43:20.8 & 0.560 & 75 & 0.444 & 0.068 & 6495 & 6049 & 0\\
GMRTEN1~J160323.1$+$543737 & 16:03:23.13 & $+$54:37:37.3 & 1.920 & 74 & 2.256 & 0.119 & 6564 & 3421 & 0\\
GMRTEN1~J160324.0$+$550618 & 16:03:24.05 & $+$55:06:18.2 & 0.635 & 84 & 0.495 & 0.079 & 6527 & 4568 & 0\\
GMRTEN1~J160326.0$+$540817 & 16:03:26.00 & $+$54:08:17.1 & 0.595 & 94 & 0.486 & 0.081 & 6579 & 2248 & 0\\
GMRTEN1~J160327.7$+$552647 & 16:03:27.78 & $+$55:26:47.3 & 3.240 & 84 & 3.867 & 0.149 & 6484 & 5387 & 0\\
GMRTEN1~J160327.9$+$543326 & 16:03:27.99 & $+$54:33:26.0 & 0.452 & 75 & 0.338 & 0.064 & 6540 & 3253 & 0\\
GMRTEN1~J160329.5$+$540705 & 16:03:29.55 & $+$54:07:05.7 & 0.869 & 131 & 0.681 & 0.117 & 6559 & 2200 & 0\\
GMRTEN1~J160330.8$+$542454 & 16:03:30.83 & $+$54:24:54.3 & 0.483 & 78 & 0.837 & 0.097 & 6533 & 2912 & 0\\
GMRTEN1~J160331.1$+$554327 & 16:03:31.14 & $+$55:43:27.2 & 0.473 & 76 & 0.555 & 0.071 & 6447 & 6052 & 0\\
GMRTEN1~J160331.3$+$545000 & 16:03:31.39 & $+$54:50:00.7 & 0.408 & 67 & 0.322 & 0.055 & 6503 & 3916 & 0\\
GMRTEN1~J160332.3$+$553000 & 16:03:32.38 & $+$55:30:00.2 & 0.508 & 69 & 0.536 & 0.076 & 6454 & 5514 & 0\\
GMRTEN1~J160332.6$+$554622 & 16:03:32.66 & $+$55:46:22.6 & 3.212 & 76 & 4.152 & 0.153 & 6435 & 6169 & 0\\
GMRTEN1~J160332.9$+$541746 & 16:03:32.90 & $+$54:17:46.8 & 0.474 & 73 & 0.381 & 0.060 & 6528 & 2627 & 0\\
GMRTEN1~J160333.1$+$542914 & 16:03:33.11 & $+$54:29:14.2 & 2.377 & 73 & 9.697 & 0.226 & 6515 & 3085 & 3\\
GMRTEN1~J160333.6$+$552623 & 16:03:33.69 & $+$55:26:23.7 & 0.773 & 78 & 0.878 & 0.101 & 6451 & 5370 & 0\\
GMRTEN1~J160333.7$+$540540 & 16:03:33.70 & $+$54:05:40.6 & 7.541 & 294 & 9.799 & 0.552 & 6536 & 2142 & 0\\
GMRTEN1~J160334.6$+$542900 & 16:03:34.68 & $+$54:29:00.2 & 2.326 & 80 & 4.235 & 0.167 & 6506 & 3075 & 3\\
GMRTEN1~J160335.1$+$551534 & 16:03:35.19 & $+$55:15:34.1 & 0.462 & 74 & 0.333 & 0.066 & 6454 & 4937 & 0\\
GMRTEN1~J160335.2$+$555419 & 16:03:35.24 & $+$55:54:19.5 & 0.762 & 105 & 0.984 & 0.108 & 6412 & 6486 & 0\\
GMRTEN1~J160335.2$+$540515 & 16:03:35.25 & $+$54:05:15.3 & 38.494 & 383 & 50.351 & 0.961 & 6528 & 2125 & 0\\
GMRTEN1~J160336.5$+$555147 & 16:03:36.55 & $+$55:51:47.5 & 0.500 & 79 & 0.662 & 0.082 & 6408 & 6385 & 0\\
GMRTEN1~J160336.9$+$544120 & 16:03:36.90 & $+$54:41:20.7 & 0.532 & 68 & 0.853 & 0.092 & 6480 & 3568 & 0\\
GMRTEN1~J160337.6$+$545944 & 16:03:37.64 & $+$54:59:44.2 & 0.547 & 80 & 0.530 & 0.077 & 6457 & 4303 & 0\\
GMRTEN1~J160338.7$+$554348 & 16:03:38.77 & $+$55:43:48.1 & 0.771 & 80 & 1.743 & 0.135 & 6404 & 6065 & 0\\
GMRTEN1~J160339.0$+$545943 & 16:03:39.05 & $+$54:59:43.8 & 0.500 & 80 & 0.415 & 0.065 & 6449 & 4303 & 0\\
GMRTEN1~J160339.3$+$551352 & 16:03:39.37 & $+$55:13:52.0 & 0.516 & 78 & 0.414 & 0.063 & 6432 & 4868 & 0\\
GMRTEN1~J160339.6$+$542953 & 16:03:39.68 & $+$54:29:53.2 & 0.701 & 66 & 0.647 & 0.078 & 6476 & 3110 & 0\\
GMRTEN1~J160339.7$+$550600 & 16:03:39.75 & $+$55:06:00.4 & 0.584 & 80 & 0.418 & 0.069 & 6438 & 4554 & 0\\
GMRTEN1~J160340.1$+$545127 & 16:03:40.10 & $+$54:51:27.4 & 0.711 & 79 & 0.639 & 0.079 & 6451 & 3972 & 0\\
GMRTEN1~J160340.1$+$550544 & 16:03:40.15 & $+$55:05:44.9 & 0.495 & 81 & 0.392 & 0.069 & 6436 & 4543 & 0\\
GMRTEN1~J160340.8$+$554325 & 16:03:40.83 & $+$55:43:25.8 & 1.447 & 81 & 4.633 & 0.210 & 6392 & 6050 & 0\\
GMRTEN1~J160341.2$+$552611 & 16:03:41.27 & $+$55:26:11.8 & 0.475 & 68 & 0.460 & 0.065 & 6408 & 5361 & 0\\
GMRTEN1~J160341.5$+$552205 & 16:03:41.59 & $+$55:22:05.4 & 0.638 & 69 & 0.614 & 0.080 & 6411 & 5197 & 0\\
GMRTEN1~J160341.6$+$553203 & 16:03:41.66 & $+$55:32:03.4 & 0.462 & 69 & 0.365 & 0.062 & 6400 & 5595 & 0\\
GMRTEN1~J160343.1$+$540324 & 16:03:43.13 & $+$54:03:24.6 & 0.908 & 126 & 0.556 & 0.092 & 6483 & 2050 & 0\\
GMRTEN1~J160344.8$+$540320 & 16:03:44.84 & $+$54:03:20.4 & 1.168 & 118 & 0.856 & 0.110 & 6473 & 2047 & 0\\
GMRTEN1~J160345.8$+$553021 & 16:03:45.84 & $+$55:30:21.4 & 0.381 & 60 & 0.370 & 0.065 & 6378 & 5527 & 0\\
GMRTEN1~J160345.8$+$554238 & 16:03:45.88 & $+$55:42:38.3 & 0.653 & 76 & 2.566 & 0.156 & 6365 & 6018 & 0\\
GMRTEN1~J160346.5$+$552855 & 16:03:46.56 & $+$55:28:55.8 & 0.577 & 73 & 0.510 & 0.076 & 6375 & 5469 & 0\\
GMRTEN1~J160346.6$+$550826 & 16:03:46.62 & $+$55:08:26.0 & 0.525 & 70 & 0.416 & 0.065 & 6396 & 4650 & 0\\
GMRTEN1~J160348.3$+$542626 & 16:03:48.37 & $+$54:26:26.4 & 2.328 & 80 & 2.660 & 0.133 & 6429 & 2970 & 0\\
GMRTEN1~J160348.6$+$550124 & 16:03:48.61 & $+$55:01:24.3 & 0.477 & 78 & 0.257 & 0.050 & 6392 & 4368 & 0\\
GMRTEN1~J160349.2$+$554243 & 16:03:49.21 & $+$55:42:43.9 & 3.349 & 85 & 4.224 & 0.163 & 6346 & 6021 & 0\\
GMRTEN1~J160350.0$+$545634 & 16:03:50.06 & $+$54:56:34.6 & 0.435 & 67 & 0.371 & 0.058 & 6389 & 4175 & 0\\
GMRTEN1~J160350.4$+$550717 & 16:03:50.43 & $+$55:07:17.2 & 0.597 & 78 & 0.626 & 0.083 & 6376 & 4603 & 0\\
GMRTEN1~J160350.5$+$541302 & 16:03:50.51 & $+$54:13:02.7 & 0.770 & 80 & 0.535 & 0.068 & 6430 & 2435 & 0\\
GMRTEN1~J160351.1$+$555644 & 16:03:51.13 & $+$55:56:44.4 & 3.996 & 120 & 7.391 & 0.287 & 6321 & 6581 & 0\\
GMRTEN1~J160351.2$+$552346 & 16:03:51.24 & $+$55:23:46.1 & 0.519 & 80 & 0.640 & 0.072 & 6354 & 5262 & 0\\
GMRTEN1~J160351.3$+$540609 & 16:03:51.38 & $+$54:06:09.9 & 0.951 & 110 & 0.843 & 0.110 & 6432 & 2159 & 0\\
GMRTEN1~J160352.6$+$550012 & 16:03:52.62 & $+$55:00:12.8 & 0.510 & 71 & 0.592 & 0.076 & 6370 & 4320 & 0\\
GMRTEN1~J160352.8$+$542943 & 16:03:52.80 & $+$54:29:43.5 & 2.258 & 82 & 2.422 & 0.124 & 6400 & 3101 & 0\\
GMRTEN1~J160353.2$+$550309 & 16:03:53.29 & $+$55:03:09.9 & 0.570 & 74 & 0.500 & 0.072 & 6363 & 4438 & 0\\
GMRTEN1~J160355.5$+$553844 & 16:03:55.52 & $+$55:38:44.9 & 0.463 & 73 & 0.242 & 0.050 & 6314 & 5861 & 0\\
GMRTEN1~J160356.2$+$550126 & 16:03:56.20 & $+$55:01:26.7 & 0.472 & 77 & 0.380 & 0.063 & 6348 & 4369 & 0\\
GMRTEN1~J160356.2$+$550415 & 16:03:56.25 & $+$55:04:15.3 & 0.506 & 83 & 0.451 & 0.068 & 6345 & 4481 & 0\\
\hline
  \end{tabular}
\end{table*}

\subsection{Comparison with other surveys}
\begin{figure}
  \includegraphics[width=8cm]{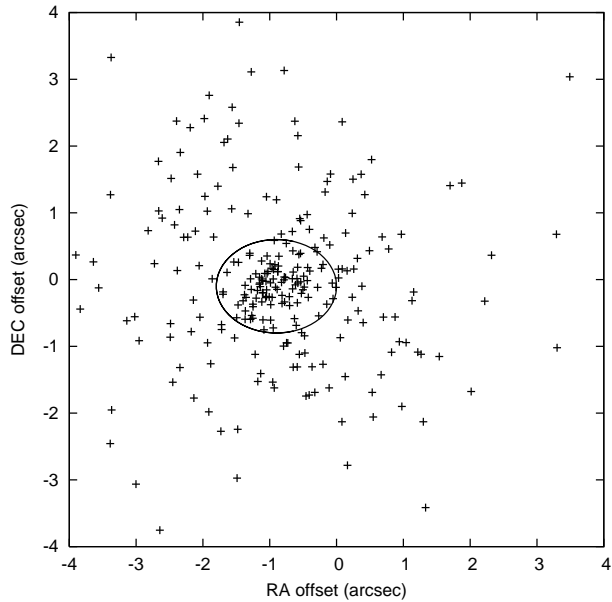}
  \caption{Source positions in the GMRT catalogue relative to the
  positions found in the VLA catalogue of \citet{Ciliegi99}, for
  unique matches within 6~arcsec.  Offsets are distributed in a
  Gaussian fashion in RA and DEC, and the ellipse corresponding to
  1$\sigma$ for the distribution is shown.}
  \label{fig:deltaRADEC}
\end{figure}

In order to test the positional accuracy of our catalogue, we paired
it with the 393 objects in the VLA 1.4~GHz source catalogue of
\citet{Ciliegi99}, using a pairing radius of 6~arcsec.  The VLA survey
covers only $\sim25$~per~cent of our 610~MHz observations.
Fig.~\ref{fig:deltaRADEC} shows the position offset of the 263 matched
sources compared with their VLA counterparts -- the distribution of
offsets is approximately Gaussian, with mean offset in Right Ascension
of $-0.9$~arcsec, and $-0.1$~arcsec in Declination.  The standard
deviations of the distribution are 0.9 and 0.7~arcsec respectively.

We repeated this analysis with data from the FIRST survey
\citep{Becker95}.  The whole of our 610~MHz survey region is covered
by FIRST, and even with the reduced sensitivity of FIRST
($\sim$150~$\mu$Jy noise), 504 pairs are found within 6~arcsec.  We
again find Gaussian-distributed position errors, with RA offset of
$-0.1$~arcsec, standard deviation 0.4~arcsec and DEC offset of
$-0.1$~arcsec, standard deviation 0.6~arcsec.  We have not corrected
the positions given in our catalogue, since it agrees closely with
FIRST.

\begin{figure}
  \includegraphics[width=8cm]{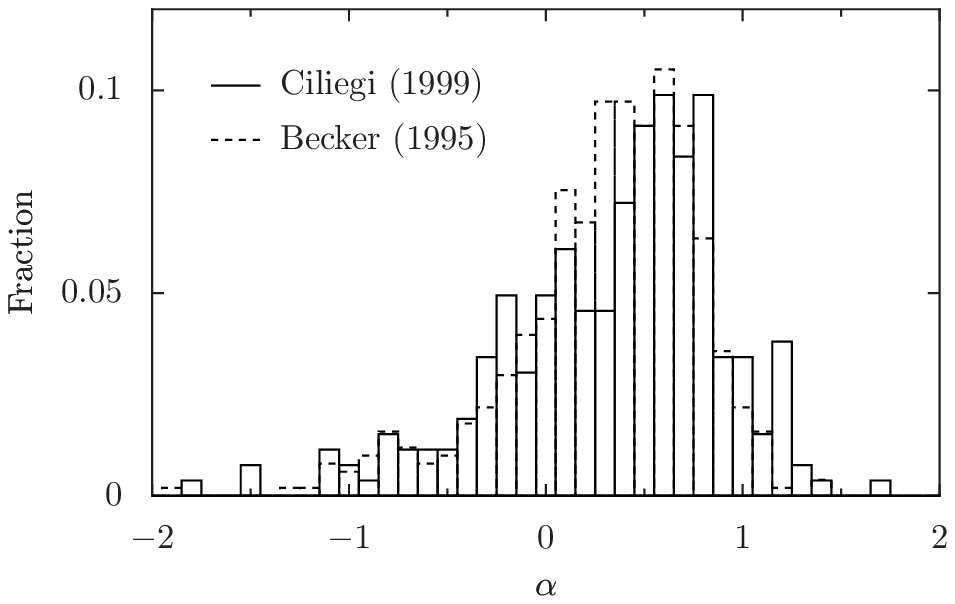}
  \caption{Radio spectral index $\alpha$ between 610~MHz and 1.4~GHz,
  for sources in the VLA ELAIS-N1 catalogue of Ciliegi et al.\ (solid
  lines) and in the FIRST catalogue (dashed lines).}
  \label{fig:alpha}
\end{figure}

The spectral index distribution of the matched sources from both
surveys is shown in Fig.~\ref{fig:alpha}, using the integrated flux
density measurements.  The distribution peaks around $\alpha=0.7$,
where $\alpha$ is defined such that the flux density $S$ scales with
frequency $\nu$ as $S = S_{0}\nu^{-\alpha}$.  

\begin{figure}
  \includegraphics[width=8cm]{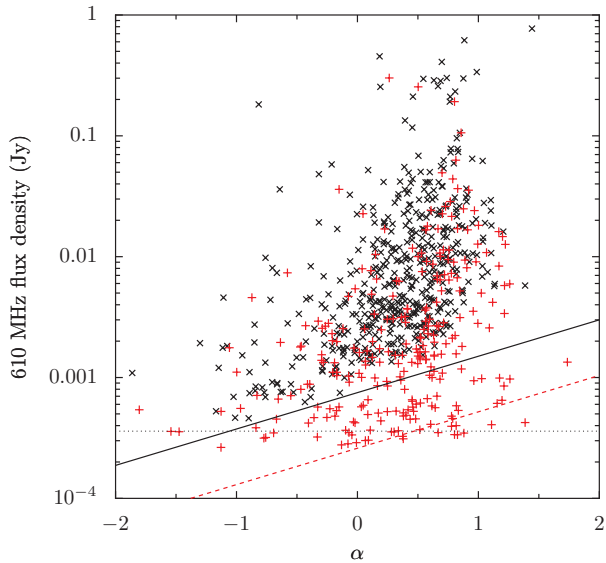}
  \caption{The variation in spectral index $\alpha$ with 610~MHz flux
  density.  Black diagonal crosses represent sources in the GMRT and
  FIRST catalogues, with the solid black line showing the limiting
  spectral index that could be detected, given the respective
  sensitivity levels.  Red upright crosses represent sources in the
  corresponding GMRT and Ciliegi et al.\ catalogues, with the dashed
  red line showing the limit on $\alpha$.  The 610~MHz flux density
  limit is shown by the dotted black line.}
  \label{fig:AlphaFlux}
\end{figure}

Fig.~\ref{fig:AlphaFlux} shows the spectral index distribution for all
sources with matches in the GMRT and FIRST catalogues (black diagonal
crosses), and sources in the GMRT and \citet{Ciliegi99} catalogues
(red upright crosses).  There are significant biases in
Fig.~\ref{fig:AlphaFlux}, due to the varying sensitivity levels of the
three surveys.  In the region covered by \citet{Ciliegi99}, the
610~MHz completeness level is 360~$\mu$Jy, shown by the black dotted
line.  The limiting spectral indices for sources at the sensitivity
levels of the 1.4~GHz surveys are shown (FIRST -- solid black line,
Ciliegi -- dashed red line).  In order to look for variations in the
source population, we calculate the mean and median spectral indices
for sources with detections in the Ciliegi catalogue, with 610~MHz
flux density between 500~$\mu$Jy and 1~mJy -- the point at which the
turnover in source counts becomes visible (see
Section~\ref{sec:sourcecounts}) -- and above 1~mJy.  The mean values
of $\alpha$ are $0.22\pm0.09$ and $0.45\pm0.04$ respectively, and the
median values are $0.36\pm0.12$ and $0.56\pm0.04$.  There are 48 and
168 sources in the two flux density bins.  The bias against
steep-spectrum sources at low flux densities (which is visible in
Fig.~\ref{fig:AlphaFlux}) means that, for a source with 610~MHz flux
density of 500~$\mu$Jy, the largest value of $\alpha$ that would be
detectable is $\sim$0.8 and so this apparent flattening at fainter
flux densities may simply be due to sample bias.  However,
\citet{Bondi07} also find significantly flatter spectral indices for
fainter radio sources, again comparing 610~MHz and 1.4~GHz data, and
attribute this to the emergence of a population of low-luminosity AGNs
-- see Section~\ref{sec:sourcecounts} for more details.

\section{Differential source counts}
\label{sec:sourcecounts}

\begin{figure}
  \includegraphics[width=8cm]{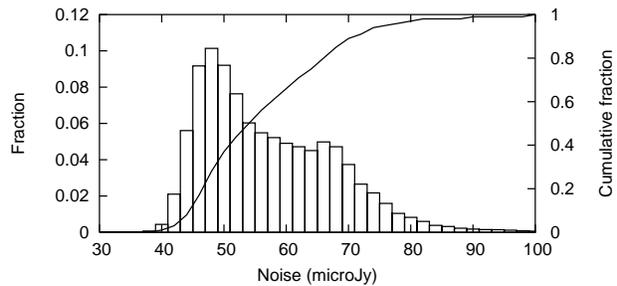}
  \caption{The fraction of pixels in the $70\times70$~arcmin$^{2}$
  region of the SExtractor rms map with each noise level (boxes, and
  left-hand axis), along with the cumulative fraction of pixels with
  noise below a particular value (solid line, and right-hand axis).}
  \label{fig:PixelValues}
\end{figure}

We derived source counts for the ELAIS-N1 field by considering two
separate regions of the map.  In order to study the 610~MHz source
counts at low flux densities, we selected a $70\times70$~arcmin$^{2}$
region (shown in Fig.~\ref{fig:EN1sample}), which was free from
spurious sources above the local $6\sigma$ value.  The region comes
from the deeper section of the survey field.
Fig.~\ref{fig:PixelValues} shows the distribution of pixel noise
levels (taken from the SExtractor rms map), along with the cumulative
fraction of pixels with noise below a given flux density.

We constructed source counts by binning our sources by their
integrated flux density, with bins ranging between 270~$\mu$Jy and
10~mJy.  The lower limit was selected as the $6\sigma$ flux density
corresponding to a noise of 45~$\mu$Jy, close to the peak in noise
distribution seen in Fig.~\ref{fig:PixelValues}.  We selected a
relatively low upper limit of 10~mJy, since our region was
specifically chosen for its lack of brighter sources (and the errors
resulting from these sources).

The source counts were corrected for the fraction of the image over
which they could be detected, taking into account the increase in
noise near the bright sources, and for the resolution bias inherent in
constructing flux density counts from a catalogue with a
signal-to-noise ratio based around the peak brightness of a source.
We use the resolution corrections of \citet{Moss07}, who found that
$\sim$7~per~cent of sources below 360~$\mu$Jy and 3~per~cent of
sources between 360 and 1000~$\mu$Jy were missed from a similar
resolution GMRT 610~MHz survey of the {\it XMM-Newton/Chandra} survey
field.  No correction for Eddington bias has been made, since
\citet{Moss07} find that it only affects their faintest flux density
bin, increasing it by approximately 20~per~cent.

The differential source count ${\rm d}N/{\rm d}S$ was calculated by dividing
$N_{\rm c}$, the corrected number of sources in each bin, by $A \Delta
S$, where $A$ is the total area of the image in steradians, and
$\Delta S$ is the width of the flux bin in Jy, i.e.

\begin{equation}
  \frac{{\rm d}N}{{\rm d}S} = \frac{N_{\rm c}}{A \Delta S}
  \label{eq:dNdS}
\end{equation}

We calculated the differential source count for sources with flux
densities above 10~mJy by considering the central $\sim4$~deg$^{2}$ of
the map.  This region has several bright sources with residual
sidelobes, but inspection of these residuals found them to have
typical flux densities of below 1~mJy, which will therefore not affect
the bright source counts.  No corrections for resolution bias or
Eddington bias have been made for these brighter sources.

\begin{figure*}
  \includegraphics[width=15cm]{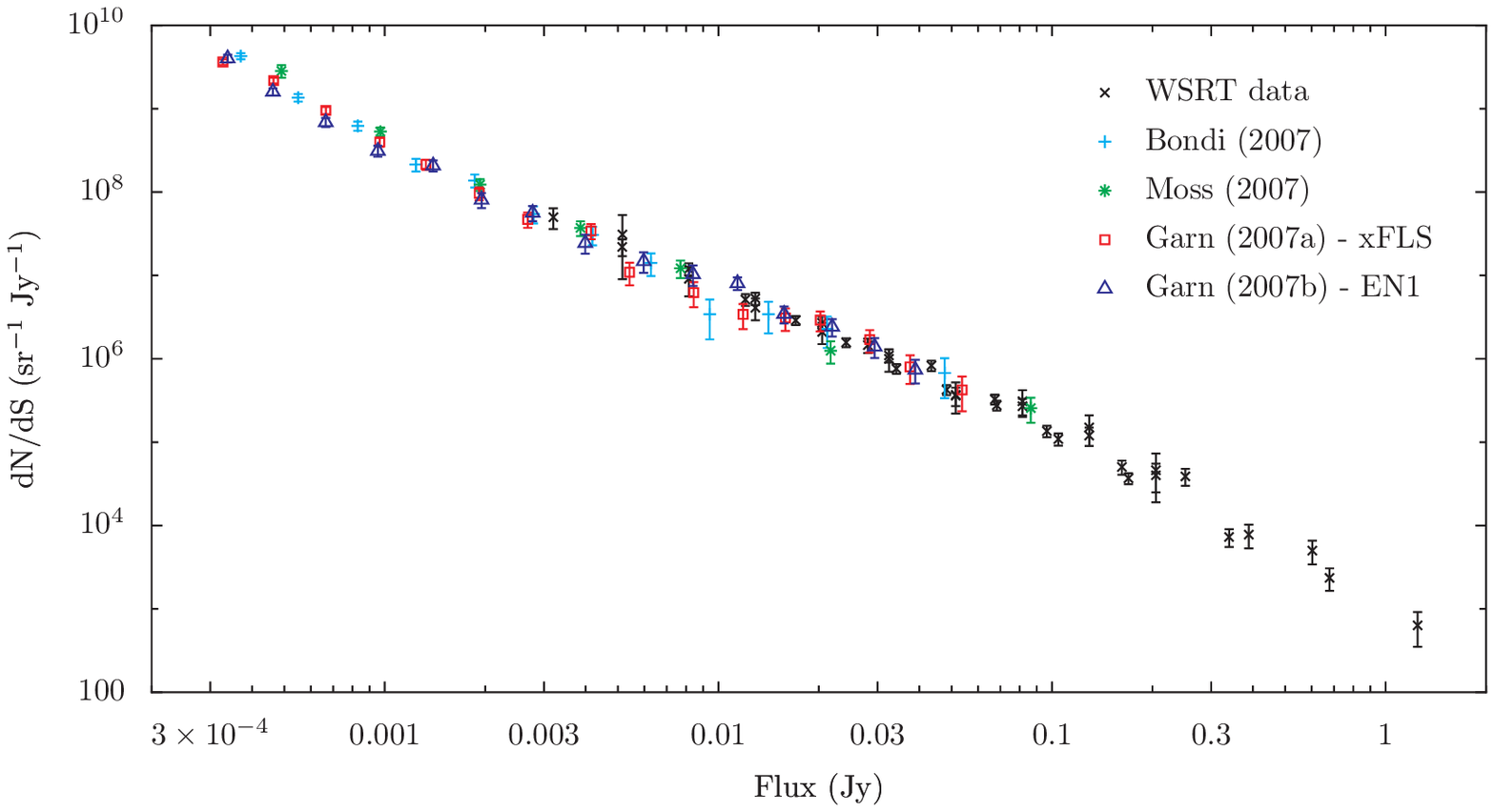}
  \caption{Differential source counts $\frac{{\rm d}N}{{\rm d}S}$ at
  610~MHz from surveys taken with the WSRT
  \citep{Valentijn77,Katgert79,Valentijn80,KatMerk85} and GMRT
  \citep{Bondi07,Moss07,Garn07}}
  \label{fig:dNdS}
\end{figure*}

Table~\ref{tab:dNdS} gives the source counts, mean flux density
$\langle S \rangle$ of sources in each bin, ${\rm d}N/{\rm d}S$ and
${\rm d}N/{\rm d}S$ normalised by $\langle S \rangle^{2.5}$, the value
expected from a static Euclidean universe.  ${\rm d}N/{\rm d}S$ is
plotted in Fig.~\ref{fig:dNdS}, along with 610~MHz source counts from
further GMRT surveys of the {\it XMM-Newton/Chandra} survey field
\citep{Moss07}, the VVDS-VLA Deep Field \citep{Bondi07}, and our
previous survey of the xFLS field \citep{Garn07}.  A series of
Westerbork Synthesis Radio Telescope (WSRT) surveys at 610~MHz
\citep[by][]{Valentijn77,Katgert79,Valentijn80,KatMerk85} are also
shown for comparison.

\begin{table}
  \label{tab:dNdS}
  \caption{610~MHz differential source counts for the ELAIS-N1 survey.}
  \begin{tabular}{cccccc} 
  \hline Flux Bin & $\langle S \rangle$ & $N$ & $N_{\rm c}$ & ${\rm
d}N/{\rm d}S$ & ${\rm d}N/{\rm d}S~\langle S \rangle^{2.5}$\\ (mJy) &
(mJy) & & & (sr$^{-1}$~Jy$^{-1}$) & (sr$^{-1}$~Jy$^{1.5}$)\\ \hline
0.270 -- 0.387 & 0.380 & 89 & 195.9 & $4.0\pm0.4\times10^{9}$ &
$8.44\pm0.89$\\ 0.387 -- 0.556 & 0.462 & 100 & 111.9 &
$1.6\pm0.2\times10^{9}$ & $7.34\pm0.73$\\ 0.556 -- 0.798 & 0.665 & 67
& 69.2 & $6.9\pm0.8\times10^{8}$ & $7.87\pm0.96$\\ 0.798 -- 1.145 &
0.953 & 45 & 45.0 & $3.1\pm0.5\times10^{8}$ & $8.77\pm1.31$\\ 1.145 --
1.643 & 1.396 & 43 & 43.0 & $2.1\pm0.3\times10^{8}$ & $15.2\pm2.31$\\
1.643 -- 2.358 & 1.950 & 24 & 24.0 & $8.1\pm1.6\times10^{7}$ &
$13.6\pm2.77$\\ 2.358 -- 3.384 & 2.775 & 24 & 24.0 &
$5.6\pm1.2\times10^{7}$ & $22.9\pm4.67$\\ 3.384 -- 4.856 & 3.989 & 15
& 15.0 & $2.4\pm0.6\times10^{7}$ & $24.7\pm6.37$\\ 4.856 -- 6.968 &
5.971 & 13 & 13.0 & $1.5\pm0.4\times10^{7}$ & $40.9\pm11.3$\\ 6.968 --
10.00 & 8.394 & 13 & 13.0 & $1.0\pm0.3\times10^{7}$ & $66.7\pm18.5$\\
10.00 -- 13.49 & 11.41 & 33 & 22.0 & $8.1\pm1.4\times10^{6}$ &
$112\pm19.6$\\ 13.49 -- 18.20 & 15.73 & 19 & 19.0 &
$3.4\pm0.8\times10^{6}$ & $107\pm24.6$\\ 18.20 -- 24.56 & 21.93 & 18 &
18.0 & $2.4\pm0.6\times10^{6}$ & $173\pm40.6$\\ 24.56 -- 33.14 & 29.41
& 14 & 14.0 & $1.4\pm0.4\times10^{6}$ & $207\pm55.3$\\ 33.14 -- 44.72
& 38.92 & 10 & 10.0 & $7.4\pm0.2\times10^{5}$ & $221\pm69.8$\\ 44.72
-- 60.34 & 53.74 & 2 & 2.0 & $1.1\pm0.1\times10^{5}$ & $73.3\pm51.9$\\
60.34 -- 81.41 & 68.60 & 3 & 3.0 & $1.2\pm0.7\times10^{5}$ &
$150.1\pm86.7$\\ 81.41 -- 109.8 & 00.00 & 0 & 0.0 & $0.0$ & 0.0 \\
109.8 -- 148.2 & 134.0 & 1 & 1.0 & $2.2\pm2.2\times10^{4}$ &
$147\pm147$\\ 148.2 -- 200.0 & 191.9 & 1 & 1.0 &
$1.7\pm1.7\times10^{4}$ & $267\pm267$\\ \hline
\end{tabular}
\end{table}

\begin{figure*}
  \includegraphics[width=15cm]{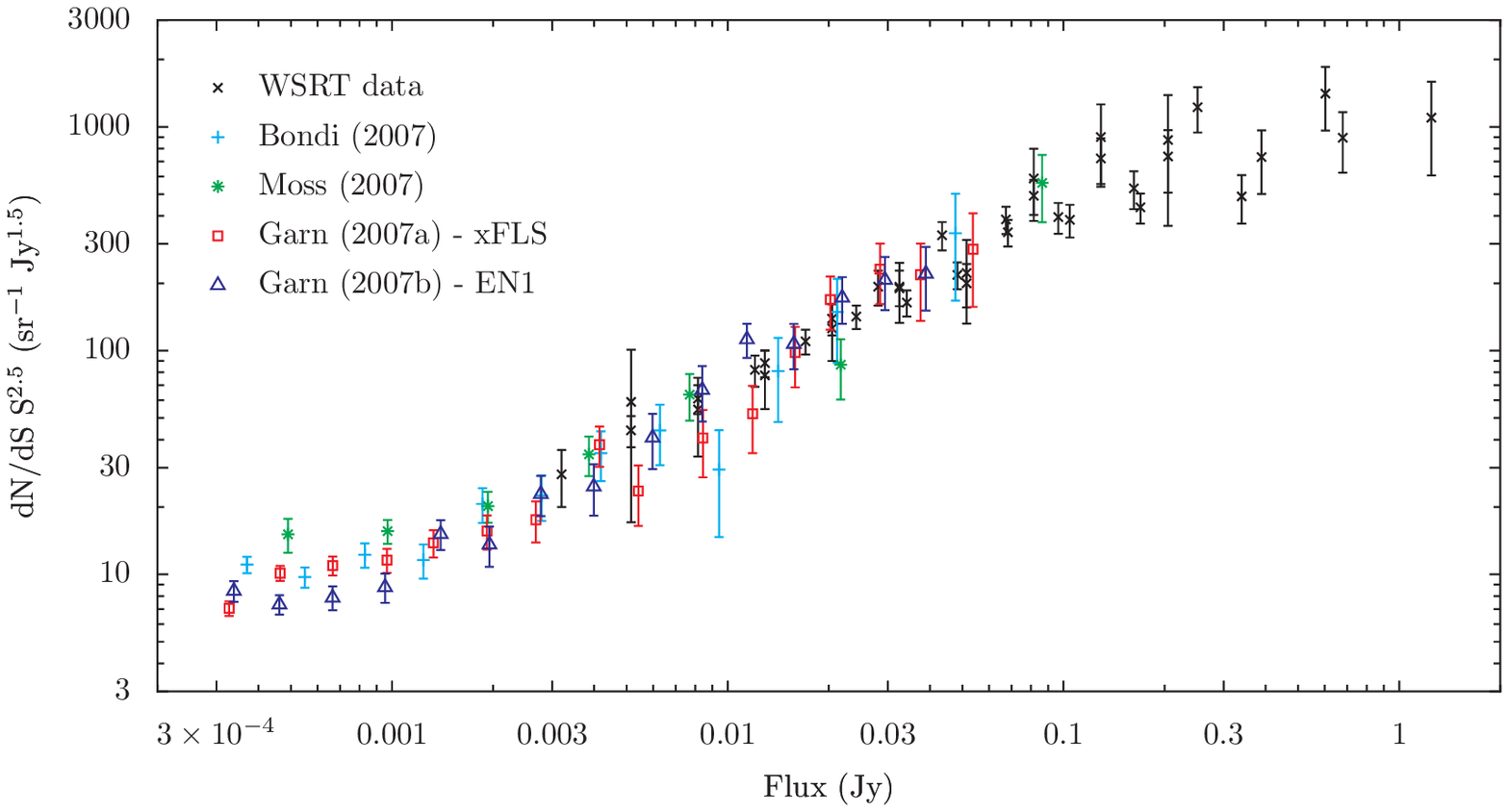}
  \caption{Differential source counts at 610~MHz
  \citep{Valentijn77,Katgert79,Valentijn80,KatMerk85,Bondi07,Moss07,Garn07},
  normalised by the value expected in a static Euclidean universe.
  The turnover in source counts is visible around 1~mJy, although
  there is considerable scatter in the various data sets below this
  flux density.}
  \label{fig:NumDist}
\end{figure*}

Fig.~\ref{fig:NumDist} shows the normalised differential source
counts.  There is good agreement in the various surveys above 1~mJy,
and all surveys show a turnover just above 1~mJy although there is
approximately a factor of 2 discrepancy between the highest and lowest
points in the value of ${\rm d}N/{\rm d}S~\langle S \rangle^{2.5}$
below this flux density. A variation in source counts has been
previously observed at 1.4~GHz \citep[see
e.g.][]{Seymour04,Huynh05,Biggs06}.  In particular, \citet{Biggs06}
and \citet{Simpson06} find a factor of 2 difference in the 1.4~GHz
source counts for different fields at $\sim$180~$\mu$Jy, equivalent to
$\sim$350~$\mu$Jy at 610~MHz for an assumed spectral index of 0.8.
This difference could be due to cosmic variance and the small numbers
of sources detected in each field at these depths, and further data is
needed to determine where the true source counts lie at these flux
density levels.

The turnover in source counts is a well-known feature of the source
counts at 1.4~GHz, and is thought to be due to the emergence of a new
population of sources, with some authors \citep[e.g.][]{Condon89,
RowanRobinson93, Hopkins98, Seymour04} attributing it to star-forming
galaxies, while others \citep{Jarvis04,Simpson06} claiming that
low-luminosity AGNs may form an important contribution to the source
counts at these levels.  The turnover occurs at 1~mJy at 1.4~GHz,
equivalent to $\sim$1.9~mJy at 610~MHz for a spectral index of 0.8,
which is consistent with Fig.~\ref{fig:NumDist}.

\begin{figure*}
  \includegraphics[width=15cm]{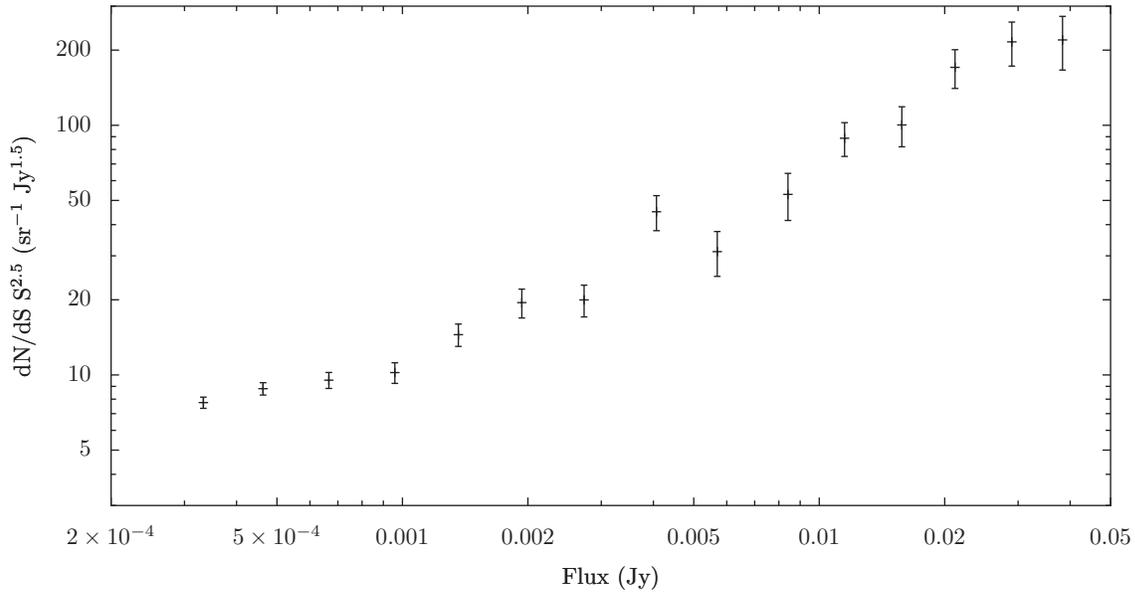}
  \caption{The normalised 610~MHz differential source counts, found by
  combining this work and the xFLS survey of \citet{Garn07}.}
  \label{fig:Overall}
\end{figure*}

In Fig.~\ref{fig:Overall} we show the differential source counts from
the combination of our ELAIS-N1 and {\it Spitzer} extragalactic First
Look Survey field -- there are more than a hundred sources in the
lowest four flux bins, leading to the most precise estimation of the
610~MHz source counts to date.

\section*{Acknowledgements}

We thank the staff of the GMRT who have made these observations
possible.  TG thanks the UK STFC for a Studentship.  The GMRT is
operated by the National Centre for Radio Astrophysics of the Tata
Institute of Fundamental Research, India. 

\bibliography{./References.bib}
\bibliographystyle{mn2e}

\label{lastpage}

\end{document}